\documentclass[reprint,prd,nofootinbib,superscriptaddress]{revtex4}
\usepackage{amsfonts}
\usepackage{amsmath}
\usepackage{amssymb}
\usepackage[english]{babel}
\usepackage{graphicx}
\usepackage{epsfig}
\usepackage{bm}
\usepackage{verbatim}
\usepackage[utf8]{inputenc}
\usepackage{booktabs}
\usepackage{multirow}
\usepackage{xcolor}
\usepackage[colorlinks=true,urlcolor=red,citecolor=red]{hyperref}
\usepackage[font=small]{caption}
\usepackage{float}
\usepackage{blindtext}
\usepackage{subfigure}
\usepackage{float}
\usepackage{cancel}
\usepackage{csquotes}
\usepackage{ulem}

\newcommand{\nn}{\nonumber}

\newcommand{\bc}{\begin{center}}
\newcommand{\ec}{\end{center}}

\newcommand{\mathsym}[1]{}

\begin{document}

\title{Phenomenological aspects of the fermion and scalar sectors of a $S_4$
flavored 3-3-1 model}
\author{A. E. C\'arcamo Hern\'andez}
\email{antonio.carcamo@usm.cl}
\affiliation{Universidad T\'{e}cnica Federico Santa Mar\'{\i}a, Casilla 110-V, Valpara\'{\i}so, Chile.}
\affiliation{Centro Científico-Tecnológico de Valparaíso, Casilla 110-V, Valparaíso, Chile.}
\affiliation{Millennium Institute for Subatomic physics at high energy frontier - SAPHIR, Fernandez Concha 700, Santiago, Chile.}
\author{Juan Marchant Gonz\'{a}lez}
\email{juan.marchant@upla.cl}
\affiliation{Laboratorio de C\'omputo de F\'isica (LCF-UPLA), Facultad de Ciencias Naturales y Exactas, Universidad de Playa Ancha, Subida Leopoldo Carvallo 270, Valpara\'iso, Chile.}
\affiliation{Departamento de Física, Universidad T\'{e}cnica Federico Santa Mar\'{\i}a. Casilla 110-V, Valpara\'{\i}so, Chile.}
\affiliation{Millennium Institute for Subatomic physics at high energy frontier - SAPHIR, Fernandez Concha 700, Santiago, Chile.}
\author{M.L. Mora-Urrutia}
\email{maria.luisa.mora.u@gmail.com}
\affiliation{Departamento de Física, Universidad T\'{e}cnica Federico Santa Mar\'{\i}a. Casilla 110-V, Valpara\'{\i}so, Chile.}
\author{Daniel Salinas-Arizmendi}
\email{daniel.salinas@usm.cl}
\affiliation{Departamento de Física, Universidad T\'{e}cnica Federico Santa Mar\'{\i}a. Casilla 110-V, Valpara\'{\i}so, Chile.}
\affiliation{Centro Científico-Tecnológico de Valparaíso, Casilla 110-V, Valparaíso, Chile.}
\date{\today }

\begin{abstract}
We proposed a viable and predictive model based on the $SU(3)_C \times SU(3)_L \times U(1)_X$ gauge symmetry, supplemented by the global $U(1)_{Lg}$ symmetry, the $S_4$ family symmetry and several auxiliary cyclic symmetries, which successfully reproduces the experimentally observed SM fermion mass and mixing pattern. The tiny active neutrino masses are generated through an inverse seesaw mechanism mediated by right-handed Majorana neutrinos. The model is consistent with the SM fermion masses and mixings and successfully accommodates the current Higgs diphoton decay rate constraints as well as the constraints arising from oblique $S$, $T$ and $U$ parameters and we studied the meson mixing due to flavor changing neutral currents mediated by heavy scalars, finding parameter space consistent with experimental constraints.
\end{abstract}

%\pacs{14.60.St, 11.30.Hv, 12.60.-i, 12.60.Cn, 12.60.Fr}
\maketitle

\section{\label{intro}Introduction}

The Standard Model of particles is a widely accepted theory describing subatomic particles' fundamental interactions. This theory has successfully explained and predicted numerous phenomena observed in particle physics experiments. Despite the great success of the Standard Model (SM), it has several unresolved problems. For example, the SM cannot explain the fermion sector's hierarchy of masses and mixings. The range of fermion mass values extends approximately 13 orders of magnitude from the light-active neutrino mass scale up to the mass of the top quark. Whereas in the lepton sector two of the mixing angles are large and one is small, the mixing angles of the quark sector are very small, thus implying that the quark mixing matrix approaches the identity matrix. These different mass and mixing patters in the quark and lepton sectors corresponds to the SM flavor puzzle, which is not explained by the SM.  %However, the mixing in the lepton sector is more considerable, known as the flavor puzzle.
 In addition to this, there are other issues not explained by the SM, such as, for example, the number of families of SM fermions as well the quantization of the electric charge. 
%cannot explain the number of families of SM fermions as well the quantization of the electric charge
%, these being some of the problems that cannot be explained through 
%the SM,
Explaining the aforementioned issues motivates to consider extensions of the SM, with enlarged particle spectrum and symmetries. %us to search for physics beyond the Standard Model (BSM).\\
%Due to these and other problems, there is a motivation to search for new physics beyond the SM, where one can find proposals such as extending the symmetry of the SM with discrete symmetries, increasing the particle content of the SM, seesaw models, etc. 
Possible SM extension options include models based on the $SU(3)_C\times SU(3)_L\times U(1)_X$ gauge group (also known as 3-3-1 models). These models have been extensively worked on in the literature \cite{Valle:1983dk,Pisano:1991ee,Frampton:1992wt,Foot:1994ym,Long:1995ctv,CarcamoHernandez:2005ka,Chang:2006aa,Hernandez:2013mcf,CarcamoHernandez:2013krw,Boucenna:2014ela,CarcamoHernandez:2014wdl,Okada:2015bxa,Hernandez:2016eod,Fonseca:2016tbn,CarcamoHernandez:2017cwi,CarcamoHernandez:2018iel,CarcamoHernandez:2019vih,CarcamoHernandez:2019iwh,CarcamoHernandez:2019lhv,CarcamoHernandez:2020ehn,Binh:2020aal,CarcamoHernandez:2021tlv,Hernandez:2021zje,Thu:2023xai,Ciftci:2022lai,Ciftci:2022lrc} because they provide an explanation of the number of fermion generations and why there are non-universal gauge assignments under the group $U(1)_X$ for left-handed quark fields (LH), which implies the cancellation of chiral anomalies when the number of the fermionic triplet and the anti-triplet of $SU(3)_L$ are equal, which occurs when the number of fermion families is a multiple of 3. Another feature of these 3-3-1 models is that the Peccei-Quinn (PQ) symmetry (a solution to the strong CP problem) is obtained, which occurs naturally, and besides that, these theories contain several sources of CP violation, and explain the quantization of the electric charge.
There are two most widely used versions of 3-3-1 models, a minimal one where the lepton components are in the same triplet representation $(\nu,l,l^C)_L$ and a variant where right-handed (RH) neutrinos are included in the same triplet $(\nu,l,\nu^C)_L$. In addition, within the framework of 3-3-1 models, research focuses on implementing radiative Seesaw mechanisms and non-renormalizable terms through a Froggen-Nilsen (FN) mechanism to explain the pattern of mass and mixing of SM fermions. It should be noted that the FN mechanism will not produce a new break scale since the flavor break scale is the same as the symmetry break scale of the 3-3-1 model.

In this work, we propose an extension of the SM through a 3-3-1 model with (RH) neutrinos, also adding a global lepton symmetry $U(1)_{Lg}$ to ensure the conservation of the lepton number, a discrete non-abelian symmetry $S_4 $ to reproduce the masses and mixing of the fermionic sector and three other auxiliary cyclic symmetries $Z_4\times Z_4^{\prime}\times Z_2$, where the $Z_4$ symmetry is introduced to obtain a texture with zeros in the entries of the up type quark mass matrix, in addition, the symmetry $Z_4^{\prime}$ together with the VEV pattern of the scalar triplets associated with the sector of charged leptons is necessary to get a diagonal charged lepton mass matrix. Finally, the $Z_2$ symmetry is necessary to obtain all the complete entries of the third column of the mass matrix of the down quarks and thus with the values of the VEVs of the scalars that participate in the Yukawa terms of this sector. Given that the charged lepton mass matrix is predicted to be diagonal, the lepton mixing entirely arises from the neutrino sector, where we will use an inverse seesaw mechanism \cite{Mohapatra:1986bd} mediated by right-handed heavy neutrinos to generate the tiny active neutrinos masses. 
The $S_4$ symmetry group is a compelling choice due to its unique properties and its ability to efficiently describe the observed pattern of fermion masses and mixing angles in the standard model. As the smallest non-abelian group with irreducible representations of doublet, triplet, and singlet, $S_4$ allows for an elegant accommodation of the three fermion families. Furthermore, its cyclic structure and spontaneous breaking provide a suitable framework for generating fermion masses through mechanisms such as the Froggatt-Nielsen mechanism \cite{Froggatt:1978nt} for charged fermions and the inverse seesaw mechanism \cite{Mohapatra:1986bd} for light active neutrinos. The application of $S_4$ has demonstrated success in describing the observed patterns of SM fermion masses and mixings 
\cite{Altarelli:2009gn,Bazzocchi:2009da,Bazzocchi:2009pv,deAdelhartToorop:2010vtu,Patel:2010hr,Morisi:2011pm,Altarelli:2012bn,Mohapatra:2012tb,BhupalDev:2012nm,deMedeirosVarzielas:2012apl,Ding:2013hpa,Ishimori:2010fs,Ding:2013eca,Hagedorn:2011un,Campos:2014zaa,Dong:2010zu,Vien:2015fhk,deAnda:2017yeb,deAnda:2018oik,CarcamoHernandez:2019eme,CarcamoHernandez:2019iwh,deMedeirosVarzielas:2019cyj,DeMedeirosVarzielas:2019xcs,Chen:2019oey,Garcia-Aguilar:2022gfw,CarcamoHernandez:2022vjk}.
%\cite{Lam:2008rs,Altarelli:2009gn,Bazzocchi:2009da,Bazzocchi:2009pv,deAdelhartToorop:2010vtu,Patel:2010hr,Morisi:2011pm,Altarelli:2012bn,Mohapatra:2012tb,BhupalDev:2012nm,deMedeirosVarzielas:2012aplDing:2013hpa,Ishimori:2010fs,Ding:2013eca,Hagedorn:2011un,Campos:2014zaa,Dong:2010zu,Vien:2015fhk,deAnda:2017yeb,deAnda:2018oik,CarcamoHernandez:2019eme,CarcamoHernandez:2019iwh,deMedeirosVarzielas:2019cyj,DeMedeirosVarzielas:2019xcs,Chen:2019oey,Garcia-Aguilar:2022gfw}.

The non-abelian symmetry $S_4$ provides a solid theoretical framework for constructing scalar fields and understanding various phenomenologies. Scalar fields, transforming according to the representations of $S_4$, allow us to determine the phenomenon of meson oscillations. These oscillations are of great interest as they provide valuable information about flavor symmetry violations in the Standard Model. In addition, the scalar mass spectrum is analyzed and discussed in detail. Its analysis allows to study the implications of our model in the decay of the Standard Model like Higgs boson into a photon pair as well as in meson oscillations. %production of photons through the decay of the Standard Model Higgs. 
Finally, besides providing new physics contribution to meson mixingz and Higgs decay into two photons, the considered model succesfully satisfies the constraints imposed by the oblique parameters, where the corresponding analysis is performed consideing the low energy effective field theory below the scale of spontaneous breaking of the $SU(3)_L\times U(1)_X\times U(1)_{L_g}$ symmetry. %at the low energy limit. 
These parameters, derived from precision measurements in electroweak physics, are essential for evaluating the consistency between the theoretical model and experimental data. The consistency of our model with the oblique parameters demonstrates its ability to reproduce experimental observations in the context of electroweak physics for energy scales below $1$ TeV.

This paper is organized as follows. The section \ref{model} presents the model and its details, such as symmetries, particle content, field assignments under the symmetry group, and describe the spontaneous symmetry breaking pattern. In the sections \ref{quark-sector} and \ref{lepton-sector}, the implications of the model in the masses and mixings in quark and lepton sectors, respectively, are discussed and analyzed. In addition, section \ref{scalar} describes the scalar potential, the resulting scalar mass spectrum and the mixing in the scalar sector. The section \ref{meson} provides an analysis and discussion of the phenomenological implications of the model in meson mixings. On the other hand, the decay rate of the Higgs to two photons is also studied in section \ref{di-photon}. In section \ref{oblique}, the contribution of the model to the oblique parameters through the masses of the new scalar fields is discussed. We state our conclusions in section \ref{conclu}.

%%%%%%%%%%%%%%%%%%%%%%%%%%%%%%%%%%%%%%%%%%%%%%%%%%%%%%%%%%%%%%%%%%%%%%%

\section{\label{model}The model}

 The model is based on the $SU(3)_C \times SU(3)_L \times U(1)_X $ gauge symmetry, supplemented by the $S_4$ family symmetry and the $Z_4 \times Z'_{4} \times Z_2$ auxiliary cyclic symmetries, whose spontaneous breaking generates the observed SM fermion mass and mixing pattern. We also introduce a global $U(1)_{L_g}$ of the generalized leptonic number $L_g$ \cite{Chang:2006aa,CarcamoHernandez:2017cwi,CarcamoHernandez:2019lhv}. That global $U(1)_{L_g}$ lepton number symmetry will be spontaneously broken down to a residual discrete lepton number symmetry $Z_2^{(L_g)}$ by a VEV of the gauge-singlet scalars $\varphi$ and $\xi$ to be introduced below. The correspoding massless Goldstone bosons, the Majoron, are phenomenologically harmless since they are gauge singlets. It is worth mentioning that under the discrete lepton number symmetry $Z_2^{(L_g)}$, the leptons are charged and the other particles are neutral, thus implying that in any interaction leptons can appear only in pair, thus, forbidding proton decay. 
  The $S_4$ symmetry is the smallest non-Abelian discrete symmetry group having five irreducible representations (irreps), explicitly, two singlets (trivial and no-trivial), one doublet and two triplets (3 and 3') \cite{Ishimori:2010au}. While the auxiliary cyclic symmetries $Z_4$, $Z'_4$ and $Z_2$ select the allowed entries of the SM fermion mass matrices that yield a viable pattern of SM fermion masses and mixings, and at the same time, the cyclic symmetries also allow a successful implementation of the inverse seesaw mechanism. The $\mathcal{G}$ chosen symmetry exhibits the following three-step spontaneous breaking:
\begin{gather}
\mathcal{G}=SU(3)_{C}\times SU(3)_{L}\times U(1)_{X}\times
U(1)_{L_{g}}\times S_{4}\times Z_{4}\times Z_{4}^{\prime }\times Z_{2} \\
\Downarrow \Lambda _{\text{int }}  \notag \\
SU(3)_{C}\times SU(3)_{L}\times U(1)_{X}\times U(1)_{L_{g}}\notag \\
\Downarrow v_{\chi }  \notag \\
SU(3)_{C}\times SU(2)_{L}\times U(1)_{Y}\times Z_2^{(L_g)}  \notag \\
\ \ \ \ \ \Downarrow v_{\eta_2},v_{\rho }  \notag \\
SU(3)_{C}\times U(1)_{Q}\times Z_2^{(L_g)}  \notag
\end{gather}%
where the different symmetry breaking scales satisfy the following hierarchy:
\begin{eqnarray}
 v=246 \text{GeV} = \sqrt{v_\rho^2+v_{\eta_2}^2} \ll v_\chi \sim \mathcal{O}\left(9.9 \right)\ \text{TeV}  \ll \Lambda_{\textbf{int}},
\end{eqnarray}
which corresponds in our model to the VEVs of the scalar fields.\\
The electric charge operator is defined \cite{Valle:1983dk,Dong:2010zu} in terms of the $SU(3)$ generators $T_3$ and $T_8$ and the identity $I_{3\times 3}$ as follows:
\begin{eqnarray}
Q=T_3+\beta T_8+ I_{3\times 3}X.
\end{eqnarray}
where for our model we choose $\beta = -1/\sqrt{3}$, and $X$ are the charge associated with gauge group $U(1)_{X}$.

The fermionic content in this model under $SU(3)_{C}\times SU(3)_{L}\times U(1)_{X}$ \cite{Valle:1983dk,Long:1995ctv,CarcamoHernandez:2017cwi}. The leptons are in triplets of flavor, in which the third component is an RH neutrino. The three generations of leptons for anomaly cancellation are:
\begin{equation}
L_{iL}=\begin{pmatrix}
\nu _{i} \\ 
l_{i} \\ 
\nu _{i}^{c} \\ 
\end{pmatrix}_{\hspace{-0.15cm}L}\sim \left(\textbf{1},\textbf{3},-1/3\right), \quad l_{iR} \sim \left(\textbf{1},\textbf{1},-1\right), \quad  N_{i R} \sim \left(\textbf{1},\textbf{1},0 \right),
\end{equation}
where $i=1,2,3$ is the family index. Here $\nu^c\equiv \nu_R^c$ is the RH neutrino and $\nu_{iL}$ are the family of neutral leptons, while $l_{iL}\ (e_L,\mu_L, \tau_L)$ is the family of charged leptons, and $N_{i R}$ are three right handed Majorana neutrinos, singlets under the $3$-$3$-$1$ group. Regarding the quark content, the first two generations are antitriplet of flavor, while the third family is a triplet of flavor; note that the third generation has different gauge content compared with the first two generations, which is required by the anomaly cancellation.
\begin{gather}
Q_{nL}=\begin{pmatrix}
d_{n} \\ 
-u_{n} \\ 
J_{n} \\ 
\end{pmatrix}_{\hspace{-0.15cm}L} \sim \left( \textbf{3},\overline{\textbf{3}},0\right),
\quad Q_{3L}=\begin{pmatrix}
u_{3} \\ 
d_{3} \\ 
T \\ 
\end{pmatrix}_{\hspace{-0.15cm}L}\sim \left( \textbf{3},\textbf{3},1/3\right), \quad n=1,2.\\
\nonumber u_{iR} \sim \left(\textbf{3},\textbf{1},2/3\right), \quad d_{iR} \sim \left(\textbf{3},\textbf{1},-1/3\right), \quad J_{nR} \sim \left(\textbf{3},\textbf{1},-1/3\right), \quad T_R \sim \left(\textbf{3},\textbf{1},2/3\right).
\end{gather}
We can observe that the $d_{iR}$ and $J_{iR}$ quarks have the same $X$ quantum number, and so are the $u_{iR}$ and $T_R$ quarks. Here $u_{i L}$ and $d_{i L}$ are the LH up and down type quarks fields in the flavor basis, respectively. Furthermore, $u_{i R}$ and $d_{i R}$ are the RH SM quarks, and $J_{n R}$ and $T_R$ are the RH exotic quarks. And the scalar sector contains four scalar triplets of flavor,
%$\rho$, $\chi$, and $\eta_n$, 
\begin{eqnarray}\label{triplet}
\nonumber \rho &=&\begin{pmatrix}
\rho _{1}^{+} \\ 
\frac{1}{\sqrt{2}}(v_{\rho }+\xi _{\rho }\pm i\zeta _{\rho }) \\ 
\rho _{3}^{+}%
\end{pmatrix}\sim \left( \textbf{1},\textbf{3},\frac{2}{3}\right) ,\quad  \chi\ =\ \begin{pmatrix}
\chi _{1}^{0} \\ 
\chi _{2}^{-} \\ 
\frac{1}{\sqrt{2}}(v_{\chi }+\xi _{\chi }\pm i\zeta _{\chi })%
\end{pmatrix}%
\sim \left( \textbf{1},\textbf{3},-\frac{1}{3}\right) ,
\\
 &&\\
\nonumber  \eta _{2} &=& \begin{pmatrix}
\frac{1}{\sqrt{2}}(v_{\eta_2}+\xi _{\eta_2}\pm i\zeta _{\eta_2}) \\ 
\eta _{22}^{-} \\ 
\eta _{32}^{0}
\end{pmatrix}\sim \left( \textbf{1},\textbf{3},-\frac{1}{3}\right), \quad  \eta_{1} \ =\ \begin{pmatrix}
\frac{1}{\sqrt{2}}(\xi _{\eta_1}\pm i\zeta _{\eta_1}) \\ 
\eta _{21}^{-} \\ 
\eta _{31}^{0}
\end{pmatrix}\sim \left( \textbf{1},\textbf{3},-\frac{1}{3}\right).
 \end{eqnarray}
where $\eta_1$ is an inert triplet scalar, on the other hand, the $SU(3)_L$ scalars $\rho$, $\chi$, and $\eta_2$ acquire the following vacuum expectation value (VEV) patterns:
\begin{equation}
\left\langle \chi \right\rangle^T= \left(0,0, v_{\chi}/\sqrt{2}\right) , \quad \left\langle \rho \right\rangle^T= \left(0, v_{\rho}/\sqrt{2}, 0\right), \quad 
\left\langle \eta_2 \right\rangle^T= \left( v_{\eta_2}/\sqrt{2},0, 0\right).
\end{equation}
In addition, some singleton scalars are introduced: $\left\lbrace   \sigma, \Theta_n, \zeta_n, \varphi , S_{k} , \Phi , \xi \right\rbrace $, $\left( k=e,\mu,\tau\right)$ where all fields transform $\left( \textbf{1},\textbf{1},0\right)$ under $SU(3)_{C}\times SU(3)_{L}\times U(1)_{X}$. 

With respect to the global leptonic symmetry defined as \cite{Chang:2006aa}:
\begin{eqnarray}
L=\frac{4}{\sqrt{3}} T_8+I_{3\times 3} L_g, 
\end{eqnarray}
where $L_g$ is a conserved charge corresponding to the $U(1)_{L_{g}}$ global symmetry, which commutes with the gauge symmetry. The difference between the $SU(3)_L$ Higgs triplet  can be explained using different charges $L_g$ the generalized lepton number. The lepton and anti-lepton are in the triplet, the leptonic number operator $L$ does not commute with gauge symmetry. \\
The choice of the $S_4$ symmetry group containing irreducible triplet, doublet, trivial singlet and non-trivial singlet representations presented in the Appendix \ref{S4}, allows us to naturally group the three charged left lepton families, and the three right-handed majonara neutrinos into $S_4$ triplets; $\left( L_{1L},L_{2L},L_{2L}\right) \sim \textbf{3}$, $N_R = \left( N_{1R},N_{2N},N_{2N}\right)\sim \textbf{3}$, while the first two families of left SM quarks, and the second and third families of right SM quarks in $S_4$ doublets;  $Q_{L} = \left( Q_{1L},Q_{2L}\right) \sim \mathbf{2}$, $U_{R}=\left( u_{2R},u_{3R}\right) \sim \mathbf{2}$ respectively, as well as the following exotic quarks; $J_{R}=\left( J_{1R},J_{2R}\right) \sim  \textbf{2} $, the remaining fermionic fields $Q_{3L}$, $T_R$, and  $l_{iR}$ as trivial singlet, $u_{1R}$ and $d_{iR}$ non-trivial singlet of $S_4$. The assignments under $S_4$, the scalar fields $S_k$, $\Phi$, and $\xi$ are grouped into triplets. In our model, the $S_k$ fields play a fundamental role in the vacuum configurations for the $S_4$ triplets leading to diagonal mass matrices for the charged leptons of the standard model. The inert field $\eta_1$ and $\eta_2$ in doublet $\eta=\left( \eta_1,\eta_2\right) \sim \textbf{2}$, there are three nontrivial singlets; $\{\sigma, \Theta_2,\ \zeta_2\} \sim \textbf{1'}$ and five trivial singlets $\{\rho, \chi, \Theta_1, \varphi, \zeta_1 \}$.

Using the particle spectrum and symmetries given in Tables \ref{Table1} and \ref{Table2}, we can write the Yukawa interactions for the quark and lepton sectors:
\begin{eqnarray}
 -\mathcal{L}_{Y}^{\left( q\right) } &=&y_{1}^{\left( T\right) }\overline{Q}%
_{3L}\chi T_{R}+y^{\left( J\right) }\left( \overline{Q}%
_{L}\chi ^{\ast }J_{R}\right) _{\mathbf{1}}+y_{3}^{\left( u\right) }%
\overline{Q}_{3L}\left( \eta U_{R}\right) _{\mathbf{1}} +y_{2}^{\left( u\right) }\varepsilon_{abc}\left( \overline{Q}_{L}^{a}\eta
^{b}\right) _{\mathbf{2}}\chi ^{c}U_{R}\frac{\sigma }{\Lambda ^{2}}\label{ec:lag-quarks}\\
\nonumber &&+y_{1}^{\left( u\right) }\varepsilon _{abc}\left( \overline{Q}_{L}^{a}\eta
^{b}\right) _{\mathbf{1^{\prime }}}\chi ^{c}u_{1R}\frac{\sigma ^{2}}{\Lambda
^{3}} +y_{13}^{\left( d\right) }\left( \overline{Q}_{L}\eta ^{\ast }\right) _{%
\mathbf{1^{\prime }}}d_{3R}\frac{\zeta _{1}}{\Lambda }+y_{12}^{\left(
d\right) }\left( \overline{Q}_{L}\eta ^{\ast }\right) _{\mathbf{1^{\prime }}%
}d_{2R}\frac{\Theta _{1}}{\Lambda }\\
\nonumber&&+y_{11}^{\left( d\right) }\left( 
\overline{Q}_{L}\eta ^{\ast }\right) _{\mathbf{1^{\prime }}}d_{1R}\frac{%
\sigma ^{2}}{\Lambda ^{2}} +y_{23}^{\left( d\right) }\left( \overline{Q}_{L}\eta ^{\ast }\right) _{%
\mathbf{1}}d_{3R}\frac{\zeta _{2}}{\Lambda }+y_{22}^{\left( d\right) }\left( 
\overline{Q}_{L}\eta ^{\ast }\right) _{\mathbf{1}}d_{2R}\frac{\Theta _{2}}{%
\Lambda } +y_{33}^{\left( d\right) }\overline{Q}_{3L}\rho d_{3R},
\end{eqnarray}
\begin{eqnarray}
-\mathcal{L}_{Y}^{\left( l\right) } &=&y_{1}^{\left( L\right) }\overline{L}_{L}\rho l_{1R}\frac{S_{e}}{\Lambda }+y_{2}^{\left( L\right) }\overline{L}_{L}\rho l_{2R}\frac{S_{\mu }}{\Lambda }+y_{3}^{\left( L\right) }\overline{L}_{L}\rho l_{3R}\frac{S_{\tau }}{\Lambda }+y_{\chi }^{\left( L\right) }\left( 
\overline{L}_{L}\chi N_{R}\right) _{\mathbf{\mathbf{1}}} \label{ec:lag-lep}  \\
\nonumber &&+y_{\nu }\varepsilon _{abc}\left( \overline{L}_{L}^{a}\rho ^{*c}\left(
L_{L}^{C}\right) ^{b}\right) _{\mathbf{3}}\frac{\Phi }{\Lambda }
+h_{1N}\left( N_{R}\overline{N_{R}^{C}}\right) _{\mathbf{\mathbf{1}}}\varphi 
\frac{\sigma ^{2}}{\Lambda ^{2}}+h_{2N}\left( N_{R}\overline{N_{R}^{C}}\right)_{\mathbf{\mathbf{3}}}\xi \frac{\sigma ^{2}}{\Lambda ^{2}}+H.c.,
\end{eqnarray}
where the $y^{(I)}$ (with $I= T, J, u, d, L$) represent the Yukawa couplings, which are dimensionless. The superscripts of these couplings indicate the specific contribution of the particles involved; $(T)$ and $(J)$, refer to the Yukawa couplings of the exotic $T$ and $J$ quarks respectively, $(u)$ and $(d)$ refer to the couplings of the quarks that give rise to the mass matrices of the up and down sectors respectively, and the index $(L)$ indicates the couplings of the charged leptonic Yukawa interactions. In the quark sector, under the $S_4$ group the couplings determine the interactions between the quark doublets $Q_L$, the quark singlets $u_{iR}$ and $d_{iR}$, the doublet $U_R$, and the scalar fields. The $h_{nN}$ terms are associated with operators involving heavy neutrinos, $h_{nN}$ refer to the interactions between the right-handed neutrinos and the scalar fields, crucial for the generation of neutrino masses. And the subscripts $\textbf{1}$, $\textbf{1}^{\prime}$, $\textbf{2}$ and $\textbf{3}$ denote the irreducible representations under the discrete symmetry group $S_4$, whose tensor products are shown in Appendix \ref{S4}.
%The parametric freedom of the scalar potential allows us to consider the following configuration of the VEV value for the $S_4$ triplets

We use the following VEV configurations (see Appendix \ref{ScalarS4doubletandtriplet}) for the $S_4$ triplets
\begin{eqnarray}
\left\langle S_{e}\right\rangle &=&v_{S_{e}}\left( 1,0,0\right) ,\hspace{1cm}%
\left\langle S_{\mu }\right\rangle =v_{S_{\mu }}\left( 0,1,0\right) ,\hspace{%
1cm}\left\langle S_{\tau }\right\rangle =v_{S_{\tau }}\left( 0,0,1\right) ,%
\hspace{1cm} \label{eq:vev-lep}\\
\left\langle \Phi \right\rangle &=&v_{\Phi }\left(1,r_1e^{i\theta},r_1e^{i\theta}\right),\hspace{1cm}\left\langle \xi \right\rangle =v_{\xi }\left(1,1,r_2\right) \label{eq:vev-lep2}
%\left\langle \Phi \right\rangle &=&v_{\Phi }\left( e^{\frac{i\pi }{2}%
%}r,1,1\right) ,\hspace{1cm}\left\langle \xi \right\rangle =v_{\xi }\left(
%1,1,1\right),
\end{eqnarray}
where $r_1$, $r_2$ and $\theta$ are free parameters.
Regarding the $S_4$ scalar doublet, we consider the following VEV pattern:
\begin{equation} \label{veveta}
\left\langle \eta \right\rangle =\frac{v_{\eta_2}}{\sqrt{2}}\left( 0,1\right),
\end{equation}
The above given VEV patterns are consistent with the scalar potential minimization conditions for a large region of parameter space. They allow to get a predictive and viable pattern of SM fermion masses and mixings as it will be shown in the next sections.

Furthermore, the $S_4$ singlet gauge singlet scalars have VEVs given by:
\begin{equation}
\left\langle \sigma \right\rangle  =  v_\sigma, \quad \left\langle \Theta_n \right\rangle =  v_{\Theta_n}, \quad \left\langle \varphi \right\rangle  = v_\varphi, \quad \left\langle \zeta_n \right\rangle= v_{\zeta_n}, \quad n=1,2.
\end{equation}

\begin{table}[]
\centering
\begin{tabular}{|c|c|c|c|c|c|c|c|c|c|c|c|c|c|c|}
\hline
& $\rho $ & $\eta$ & $\chi $ & $\sigma $ & $\Theta _{1}$ & $\Theta _{2}$ & $%
\varphi $ & $\zeta _{1}$ & $\zeta _{2}$ & $S_{1}$ & $S_2$ & $S_3$ & $\Phi $ & $\xi $ \\ 
\hline $U(1)_{L_g}$ & $-2/3$ & $-2/3$  & $4/3$ & $0$ & $0$ & $0$ & $2$ & $0$ & $0$ & $0$ &  $0$ &  $0$& $0$& $2$ \\ \hline
$S_{4}$ & $\mathbf{1}$ & $\mathbf{2}$ & $\mathbf{1}$ & $\mathbf{1^{\prime}}$
& $\mathbf{1}$ & $\mathbf{1}^{\prime }$ & $\mathbf{1}$ & $\mathbf{1}$ & $\mathbf{1}^{\prime }$ & $\mathbf{3}$ & $\textbf{3}$ &  $\textbf{3}$& $\mathbf{3}$ & $\mathbf{3}$ \\ \hline
$Z_{4}$ & $0$ & $1$ & $0$ & $-1$ & $0$ & $0$ & $0$ & $0$ & $0$ & $-1$ & $-1$ & $-1$ & $-2$ & $0$ \\ \hline
$Z_{4}^{\prime }$ & $0$ & $0$ & $0$ & $0$ & $0$ & $0$ & $0$ & $0$ & $0$ & $1$ & $1$ &  $1$ & $0$ & $0$\\ \hline
$Z_{2}$ & $-1$ & $0$ & $0$ & $0$ & $0$ & $0$ & $0$ & $-1$ & $-1$ & $0$ &$0$ &$0$ &  $1$ & $0$\\ \hline
\end{tabular}
\caption{Scalar assignments under $
U(1)_{L_{g}}\times S_{4}\times Z_{4}\times Z_{4}^{\prime }\times Z_{2}$.}
\label{Table1}
\end{table}

\begin{table}[]
\centering
\begin{tabular}{|c|c|c|c|c|c|c|c|c|c|c|c|c|c|c|}
\hline & $Q_{L}$ &  $Q_{3 L}$ & $u_{1 R}$ & $U_{R}$ &$d_{1R}$& $d_{2R}$ &$d_{3R}$  &  $T_R$ & $J_{ R}$ &    $L_L$ & $l_{1 R}$ & $l_{2 R}$ & $l_{3 R}$ & $N_R$ \\
\hline$U(1)_{L_g}$ & $2/3$& $-2/3$& $0$ & $0$ &$0$ &$0$ & $0$ & $-2$  & $2$ & $1/3$ & $1$ & $1$ & $1$& $-1$ \\
\hline$S_4$ & $\textbf{2}$ &   $\textbf{1}$ & $\textbf{1}^\prime$ & $\textbf{2}$ & $\textbf{1}^\prime$ & $\textbf{1}^\prime$ & $\textbf{1}^\prime$  &    $\textbf{1}$  &$\textbf{2}$ &   $\textbf{3}$ & $\textbf{1}$ & $\textbf{1}$ & $\textbf{1}$ &  $\textbf{3}$ \\
\hline$Z_4$ & $-1$ &$0$  & $0$ & $-1$ & $2$ & $0$ & $0$ &   $0$ & $-1$ & $-1$ & $0$ & $0$ & $0$ &  $1$   \\
\hline$Z_4'$ & $0$ & $0$ & $0$ & $0$ & $0$ & $0$ & $0$ & $0$ & $0$ & $0$ &  $-1$ & $-1$ & $-1$ & $0$   \\
\hline$Z_2$ & $0$ & $0$ & $0$ &  $0$ & $0$ & $0$ & $1$ &   $0$ &$ 0$ &  $0$ & $1$ & $1$ &  $1$ &  $0$  \\
\hline
\end{tabular}
\caption{Fermion assignments under $
U(1)_{L_{g}}\times S_{4}\times Z_{4}\times Z_{4}^{\prime }\times Z_{2}$.}
\label{Table2}
\end{table}

%%%%%%%%%%%%%%%%%%%%%%%%%%%%%%%%%%%%%%%%%%%%%%%%%%%%%%%%%%%%%%%%%%%%%%

\section{Quark masses and mixings}\label{quark-sector}

After the spontaneous symmetry breaking of the Lagrangian of Eq.\eqref{ec:lag-quarks}, we obtain the following $3\times 3$ low-scale quark mass matrices:
\begin{eqnarray}
M_{\text{U}}&=&\left( 
\begin{array}{ccc}
\frac{v_{\eta_2}v_{\chi}v_{\sigma}^2}{\Lambda^3}y_1^{(u)} & 0 & -\frac{v_{\eta_2}v_{\chi}v_{\sigma}}{\Lambda^2}y_2^{(u)} \\ 
0 & -\frac{v_{\eta_2}v_{\chi}v_{\sigma}}{\Lambda^2}y_2^{(u)} & 0 \\ 
0 & 0 & v_{\eta_2}y_3^{(u)}%
\end{array}%
\right) =\left( 
\begin{array}{ccc}
C & 0 & A \\ 
0 & A & 0 \\ 
0 & 0 & B%
\end{array}%
\right), \notag\\%\label{eq:matriz-up}
%=\frac{v}{\sqrt{2}}\left( 
%\begin{array}{ccc}
%e_{1}\lambda ^{8} & 0 & f_{1}\lambda ^{4} \\ 
%0 & f_{1}\lambda ^{4} & 0 \\ 
%0 & 0 & f_{3}%
%\end{array}%
%\right),\notag\\
M_{\text{D}}&=&\left(
\begin{array}{ccc}
-\frac{v_{\eta_2}v_{\sigma}^2}{\Lambda^2}y_{12}^{(d)} & -\frac{v_{\eta_2}v_{\Theta_1}}{\Lambda}y_{12}^{(d)} & -\frac{v_{\eta_2}v_{\zeta_1}}{\Lambda}y_{13}^{(d)} \\ 
0 & \frac{v_{\eta_2}v_{\Theta_2}}{\Lambda}y_{22}^{(d)} & \frac{v_{\eta_2}v_{\zeta_2}}{\Lambda}y_{23}^{(d)} \\ 
0 & 0 & v_{\rho}y_{33}^{(d)}%
\end{array}%
\right) = \left( 
\begin{array}{ccc}
C_{1} & A_{1} & B_{1} \\ 
0 & A_{2} & B_{2} \\ 
0 & 0 & B_{3}%
\end{array}%
\right)% =\frac{v}{\sqrt{2}} \left( 
%\begin{array}{ccc}
%c_{1}\lambda ^{7} & a_{1}\lambda ^{6} & b_{1}\lambda ^{7} \\ 
%0 & a_{2}\lambda ^{5} & b_{2}\lambda ^{5} \\ 
%0 & 0 & b_{3}\lambda ^{3}%
%\end{array}%
%\right)
, \label{MUMD}
\end{eqnarray}
%\blue{After diagonalizing the matrices \eqref{eq:matriz-up} and \eqref{MUMD}, we obtain the following eigenvalues},
%\blue{
%\begin{align}
%M_{U_{\text{diag}}}=\begin{pmatrix}
%C & 0 & 0 \\
%0 & A & 0 \\
%0 & 0 & B
%\end{pmatrix}\quad,\quad
%M_{D_{\text{diag}}}=\begin{pmatrix}
%C_1 & 0 & 0 \\
%0 & A_2 & 0 \\
%0 & 0 & B_3
%\end{pmatrix}.
%\end{align}
%}
%\marilu{where $\lambda =0.225$ is the Wolfenstein parameter and $v=246\;GeV$ is the electroweak symmetry breaking scale, with which we can characterise the hierarchy between the parameters that are defining quark mass matrix elements in Eq.(\ref{MUMD}). The masses for up-type quark are given by,}   

%\begin{equation}
%m_u=\frac{e_1 \lambda ^8 v}{\sqrt{2}}\quad,\quad m_c=\frac{f_1 \lambda ^4 v}{\sqrt{2}} \quad,\quad m_t=\frac{f_3 v}{\sqrt{2}},
%\end{equation}

%and for down-type quark, the masses are given by,
%\begin{equation}
%m_d=\frac{c_1 \lambda ^7 v}{\sqrt{2}}\quad,\quad m_s=\frac{a_2 \lambda ^5 v}{\sqrt{2}} \quad,\quad m_b=\frac{b_3 \lambda ^3 v}{\sqrt{2}}.
%\end{equation}

\begin{table}
\centering
\begin{tabular}{c|c|c} 
\hline\hline
\textbf{Observable} & \textbf{Model Value} & \textbf{Experimental Value}                        \\ 
\hline\hline
$m_u\; (\text{MeV})$       &         $2.04$      & $2.16_{-0.26}^{+0.49}$                             \\
$m_c\; (\text{GeV})$       &         $1.26$      & $1.27\pm 0.02$                                     \\
$m_t\; (\text{GeV})$       &         $172.50$      & $172.69\pm 0.30$                                   \\
$m_d\; (\text{MeV})$       &         $4.40$      & $4.67_{-0.17}^{+0.48}$                             \\
$m_s\; (\text{MeV})$       &         $93.7$      & $93.4_{-3.4}^{+8.6}$                               \\
$m_b\; (\text{GeV})$       &         $4.18$      & $4.18_{-0.02}^{+0.03}$                             \\
$\sin\theta^{(q)}_{12}$   &         $0.22530$     & $0.22500\pm 0.00067$                               \\
$\sin\theta^{(q)}_{23}$   &         $0.04332$    & $0.04182_{-0.00074}^{+0.00085}$                    \\
$\sin\theta^{(q)}_{13}$   &        $0.00390$    & $0.00369\pm 0.00011$                               \\
$J_q$               &  $2.93\times 10^{-5}$    & $\left(3.08_{-0.13}^{+0.15}\right)\times 10^{-5}$  \\
\hline\hline
\end{tabular}
\caption{The model and experimental values for observables in the quarks sector. The model value correspond to the best fit values of the quark sector observables and we use the experimental values of the quark masses at the On-Shell scale from Ref. \cite{Workman:2022ynf}.
}
\label{tab:quarks}
\end{table}

The matrices of Eq.~\eqref{MUMD} are non hermitian and we are considering the complex coupling $B_2$. However, %the square matrix is hermitian, so we consider the square matrices for the up and down quark sectors, and obtain 
the following matrices are Hermitian,
\begin{align}
M_{\text{U}}M_{\text{U}}^T= \begin{pmatrix}
A^2+C^2 & 0 & A B \\
 0 & A^2 & 0 \\
 A B & 0 & B^2
\end{pmatrix} \quad ;\quad 
M_{\text{D}}M_{\text{D}}^{\dagger}= \begin{pmatrix}
 A_1^2+B_1^2+C_1^2 & A_1 A_2+B_1 B_2 & B_1 B_3 \\
 A_1 A_2+B_1 B_2 & A_2^2+B_2^2 & B_2 B_3 \\
 B_1 B_3 & B_2 B_3 & B_3^2
\end{pmatrix}\label{eq:msqrt}
\end{align}

Considering the unitary matrices $M_{\text{U}}M_{\text{U}}^T$ (we consider the case where all entries of $M_{\text{U}}$ are real) and $M_{\text{D}}M_{\text{D}}^{\dagger}$%squared matrices
, we can fit the quark sector parameters in order to successfully reproduce the experimental values of the physical observables of the quark sector. To this end, we proceed to 
%observables by minimizing a 
minimize the following $\chi^2$ function defined as:
\begin{equation} \label{ec:funtion_error}   
\begin{aligned}
 \chi ^{2} = & \frac{\left( m_{u}^{\exp }-m_{u}^{\text{th}}\right) ^{2}}{\sigma
_{m_{u}}^{2}}+\frac{\left( m_{c}^{\exp }-m_{c}^{\text{th}}\right) ^{2}}{\sigma
_{m_{c}}^{2}}+\frac{\left( m_{t}^{\exp }-m_{t}^{\text{th}}\right) ^{2}}{\sigma
_{m_{t}}^{2}}+\frac{\left( m_{d}^{\exp }-m_{d}^{\text{th}}\right) ^{2}}{\sigma
_{m_{d}}^{2}}+\frac{\left( m_{s}^{\exp }-m_{s}^{\text{th}}\right) ^{2}}{\sigma
_{m_{s}}^{2}}\\
& +\frac{\left( m_{b}^{\exp }-m_{b}^{\text{th}}\right) ^{2}}{\sigma_{m_{b}}^{2}}
+\frac{\left( \sin{\theta}_{12}^{(q)\exp }-\sin\theta_{12}^{(q)\text{th}}\right)^{2}}{\sigma _{\sin\theta^{(q)}_{12}}^{2}}
+\frac{\left( \sin\theta_{23}^{(q)\exp }-\sin\theta _{23}^{(q)\text{th}}\right) ^{2}}{\sigma _{\sin\theta_{23}^{(q)}}^{2}}\\
& +\frac{\left( \sin\theta_{13}^{(q)\exp }-\sin\theta_{13}^{(q)\text{th}}\right) ^{2}}{%
\sigma _{\sin\theta^{(q)}_{13}}^{2}}+\frac{\left( J_q^{\exp }-J_q^{\text{th}}\right) ^{2}}{\sigma _{J_q}^{2}}\;, 
\end{aligned}
\end{equation}

where $m_i$ are the masses of the quarks ($i=u,c,t,d,s,b$), $\sin\theta^{(q)}_{ij}$ is the sine function of the quark mixing angles (with $ j , k = 1, 2, 3$) and $J_q$ is the quark Jarlskog invariant. The supra indices represent the experimental (\enquote{exp}) and theoretical (\enquote{$\text{th}$}) values, and the $\sigma$ are the experimental errors. The best-fit point of our model is shown in Table \ref{tab:quarks} together with the current experimental values, while Eq.~\eqref{eq:para} shows the benchmark point of the low energy quark sector effective parameters that allow to successfully reproduce the measured SM quark masses and CKM parameters, obtaining the following values for our best-fit point:
\begin{align}
C&=2.03\pm 4.27\times 10^{-2}\; \text{MeV} & A&= 1.26\pm 1.10\times 10^{-2}\; \text{GeV} & B&= 172.47\pm 0.19\; \text{GeV}. \notag\\
C_1&= -4.52\pm 0.10\; \text{MeV} & A_1&= 21.0\pm 0.5\; \text{MeV} & A_2&= -91.3\pm 2.1\; \text{MeV}.\label{eq:para}\\
B_1&= 14.3\pm 0.2 \; \text{MeV} & B_2&=(0.181\pm 2\times 10^{-3})e^{2.23i} \; \text{GeV} & B_3&= 4.18\pm 0.01\; \text{GeV}.\notag
\end{align}
%\vspace{5mm}
From Eq~\eqref{MUMD} for the Up quarks, we can see that the eigenvalues are $(C,A,B)$ and from Eq~\eqref{eq:para} we have that $C\simeq m_u$, $ A= m_c$ and $B\simeq m_t$, where only the parameter $A$ is equal to the mass of the \textit{charm} quark, while the other parameters are similar, but not equal to the values of the masses of the \textit{up} quark and the \textit{top} quark, this is due to the non diagonal structure of the up type quark mass matrix, which features a mixing in the 1-3 plane, as indicated by Eq~\eqref{eq:msqrt}\\ %because as we mentioned before, we are working with the square matrix of the \textit{up} quarks, which has another structure (see Eq~\eqref{eq:msqrt}) and therefore, other eigenvalues.\\
The same is true for the down quark sector, however, the structure of the matrix $M_{\text{D}}M_{\text{D}}^{\dagger}$ %its square mass matrix 
is more complicated than the one of $M_{\text{U}}M_{\text{U}}^{T}$ %square matrix of the up quarks, 
making it difficult to work with analytical expressions. For this reason, the entire analysis and diagonalization of the matrices $M_{\text{U}}M_{\text{U}}^{T}$ and $M_{\text{D}}M_{\text{D}}^{\dagger}$ have been performed %developed 
numerically, obtaining the following unitary rotation matrices for our best-fit point,
\begin{align}
R_u&=
\begin{pmatrix}
 0.999 & 0 & -0.007 \\
 0 & -1. & 0 \\
 -0.007 & 0 & -0.999
\end{pmatrix}, \\
R_d&=
\begin{pmatrix}
0.974 & 0.225 & -0.003 \\
0.225 & 0.973e^{-3.14i} & 0.043e^{-0.910i} \\
0.008e^{1.24i} & 0.043^{-2.24i} & 0.999e^{-3.14i}
\end{pmatrix}.
\end{align}

\begin{figure}[H]
\centering
\subfigure[]{\includegraphics[scale=0.35]{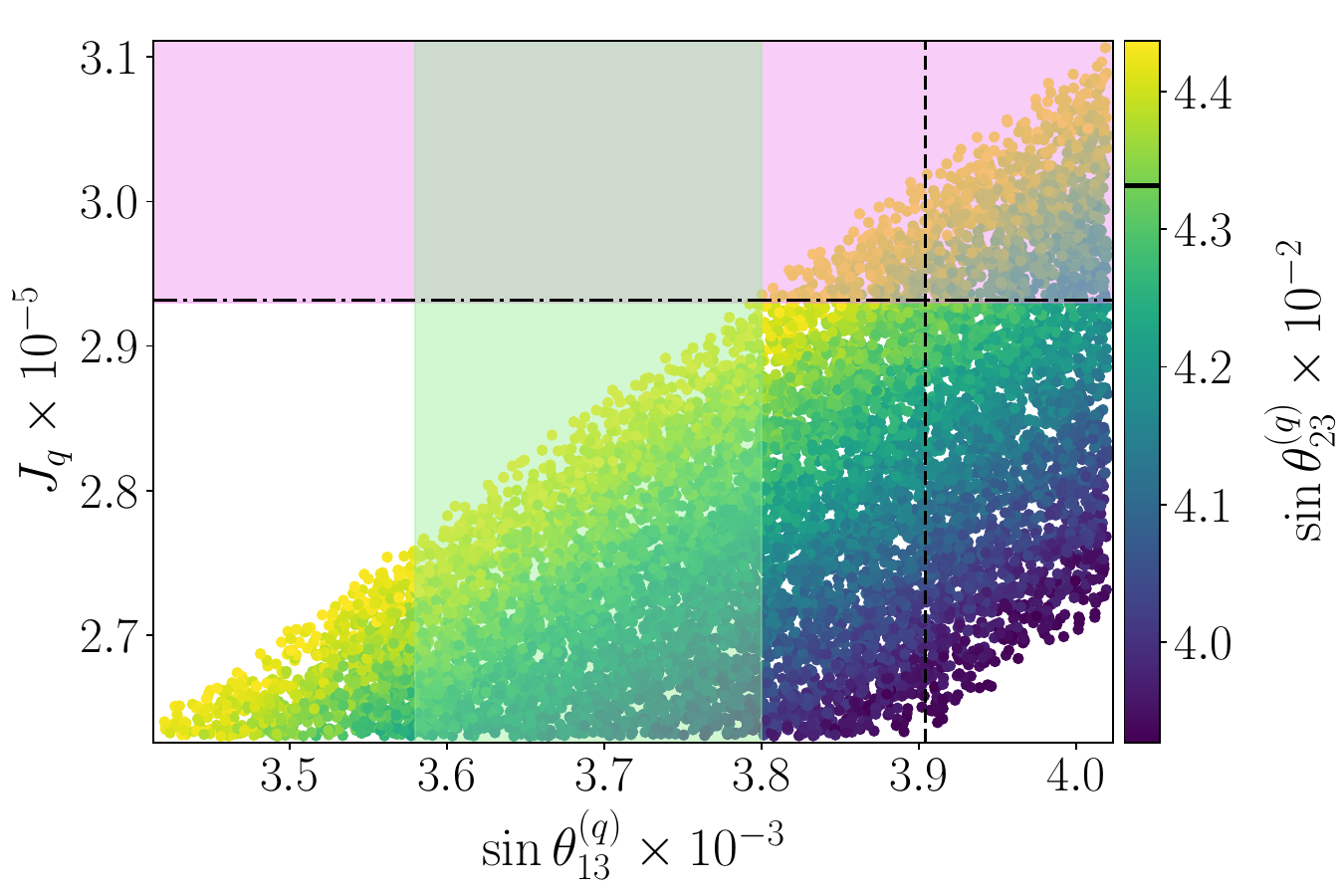}} \quad
\subfigure[]{\includegraphics[scale=0.35]{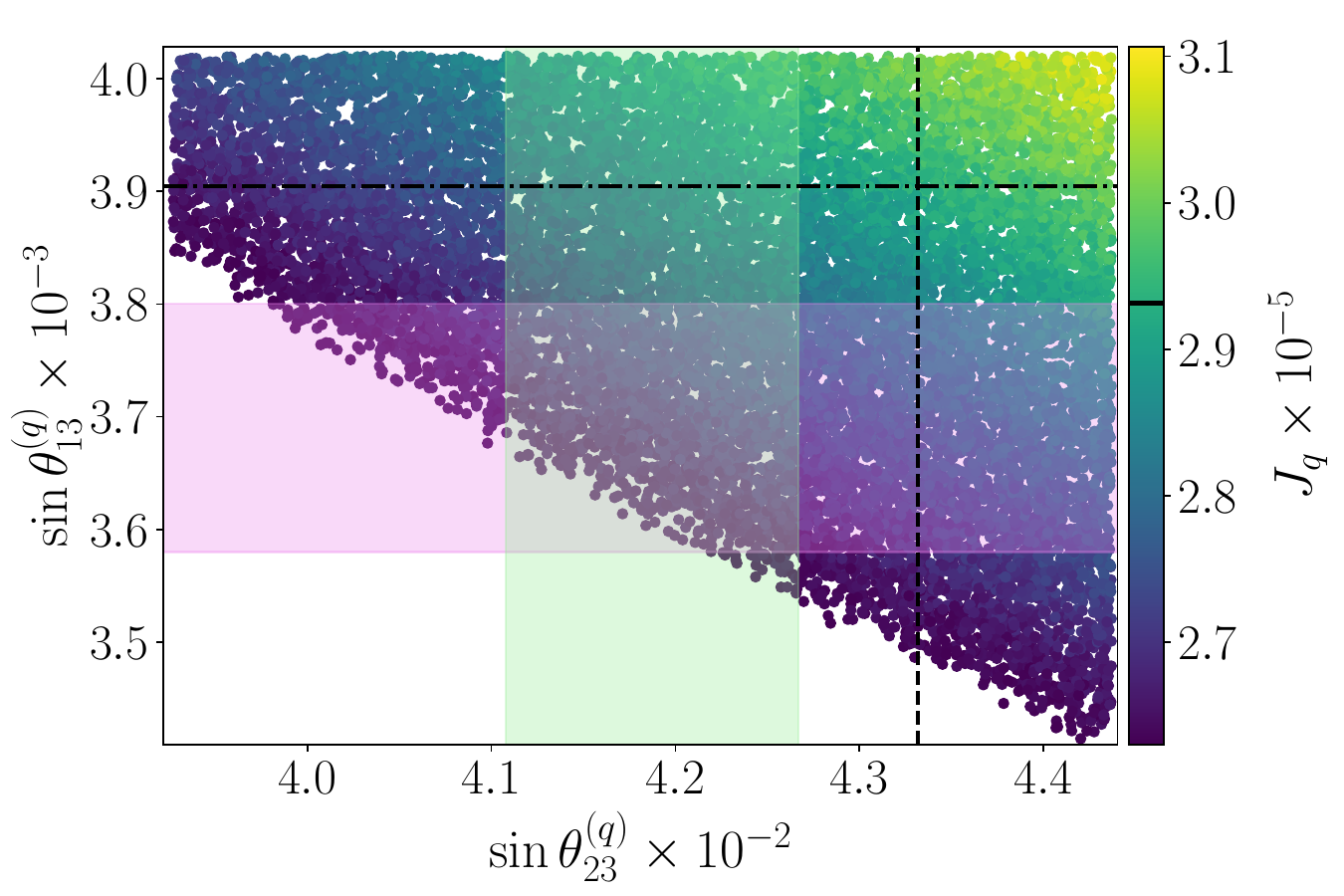}}
\caption{Correlation plot between the mixing angles of the quarks and the Jarlskog invariant obtained with our model. The green and purple bands represent the $1\sigma$ range in the experimental values, while the dotted line (black) represents the best-fit point by our model. The black line in the vertical color bar represents the best-fit point of the model for $\sin\theta_{23}^{(q)}$ and $J_q$.}
\label{fig:quarkscorr}
\end{figure}

Fig.~\ref{fig:quarkscorr} we can see the correlation plot between the quark mixing angles $\sin\theta^{(q)}_{13}$, $\sin\theta^{(q)}_{23}$ and the Jarlskog invariant.%, as well as the correlation plot between the quark mixing angles.
These correlation plots were obtained by varying the best-fit point of the quark sector parameters around $20\%$, whose values are shown in Eq.~\eqref{eq:para}. 
The dots in Fig.~\eqref{fig:quarkscorr} represent the correlation between each observable whereas the color background represents different values for  $\sin\theta_{23}^{(q)}$ (Fig.~\ref{fig:quarkscorr}a) and $J_q$ (Fig.~\ref{fig:quarkscorr}b), the vertical (green) and horizontal (purple) bars represent the $1\sigma$ range in the experimental values, while the dotted line (black) represents the value for the best-fit point of the model. The Fig.~\ref{fig:quarkscorr} a shows the correlation between $\sin\theta_{13}^{(q)}$ versus $J_q$, for different values of the $\sin\theta_{23}^{(q)}$, where the model predicts that $\sin\theta^{(q)}_{13}$ is found in the range $3.4\times 10^{-3} \lesssim \sin\theta^{(q)}_{13} \lesssim 4.0\times 10^{-3}$ in the allowed parameter space and, moreover, it increases when the Jarlskog invariant takes larger values, whose values are in the range $2.63\times 10^{-5}\lesssim J_q \lesssim 3.11\times 10^{-5}$. %Meanwhile, a similar situation occurs with $\sin\theta^{(q)}_{23}$ in \blue{Fig.~\ref{fig:quarkscorr}(b)} which is found in the range $3.9\times 10^{-2} \lesssim \sin\theta^{(q)}_{23} \lesssim 4.4\times 10^{-2}$ in the allowed parameter space and, also, it increases when the Jarlskog invariant takes larger values.
The plot, Fig.~(\ref{fig:quarkscorr}b), shows a correlation between $\sin\theta^{(q)}_{13}$ versus $\sin\theta^{(q)}_{23}$ for different values of $J_q$, in which the first variable takes on a wider range of values, with a lower limit decreasing while the upper limit remains constant, when the second one acquires larger values, where we can see that $\sin\theta^{(q)}_{23}$ is in the range $3.9\times 10^{-2} \lesssim \sin\theta^{(q)}_{23} \lesssim 4.4\times 10^{-2}$.

\begin{comment}
\begin{figure}[H]
\centering
\subfigure[]{
\includegraphics[scale=0.30]{Corrq-s13-J.png}
} \quad
\subfigure[]{
\includegraphics[scale=0.30]{Corrq-s23-J.png}
}\quad
\subfigure[]{
\includegraphics[scale=0.30]{Corrq-s23-s13.png}
}
\caption{Correlation plot between the mixing angles of the quarks and the Jarlskog invariant obtained with our model. The green and purple bands represent the $1\sigma$ range in the experimental values, while the dotted line (black) represents the best-fit point by our model.}
\label{fig:quarkscorr}
\end{figure}
\end{comment}

%%%%%%%%%%%%%%%%%%%%%%%%%%%%%%%%%%%%%%%%%%%%%%%%%%%%%%%%%%%%%%%%%%%%%%

\section{Lepton masses and mixings}\label{lepton-sector}
\subsection{Charged lepton sector}
The $Z_4^{\prime}$ charge assignments of the model fields shown in the Table \ref{Table1}, as well the VEV pattern of the $S_4$ scalar triplets $S_e$, $S_{\mu }$ and $S_{\tau}$ shown in Eq. \eqref{eq:vev-lep} imply that the charged lepton Yukawa terms of Eq.~(\ref{ec:lag-lep}), yield a diagonal charged lepton mass matrix:
\begin{equation}
M_l=\left(
\begin{array}{ccc}
m_e & 0 & 0 \\
0 & m_{\mu} & 0 \\
0 & 0 & m_{\tau}
\end{array}
\right),
\end{equation}
where the masses of the SM charged leptons are given by:
\begin{equation}
m_{e}=y_{1}^{\left( L\right) }\frac{v_{S_{e}}v_{\rho }}{\Lambda }=a_{1}\frac{%
v_{\rho }}{\Lambda },\hspace{1cm}m_{\mu }=y_{2}^{\left( L\right) }\frac{%
v_{S_{\mu }}v_{\rho }}{\Lambda }=a_{2}\frac{v_{\rho }}{\Lambda },\hspace{1cm}%
m_{\tau}=y_{3}^{\left( L\right) }\frac{v_{S_{\tau }}v_{\rho }}{\Lambda }=a_{3}%
\frac{v_{\rho }}{\Lambda }.
\end{equation}

%%%%%%%%%%%%%%%%%%%%%%%%%%%%%%%%%%%%%%%%%%%%%%%%%%%%%%%%%%%%%%%%%%%%%%
\subsection{Neutrino sector}
The neutrino Yukawa interactions of Eq. (\ref{ec:lag-lep}) give rise to the following neutrino mass terms:
\begin{equation}
-\mathcal{L}_{mass}^{\left( \nu \right) }=\frac{1}{2}\left( 
\begin{array}{ccc}
\overline{\nu _{L}^{C}} & \overline{\nu _{R}} & \overline{N_{R}}%
\end{array}%
\right) M_{\nu }\left( 
\begin{array}{c}
\nu _{L} \\ 
\nu _{R}^{C} \\ 
N_{R}^{C}%
\end{array}%
\right) +H.c,  \label{Lnu}
\end{equation}

where the neutrino mass matrix reads:
\begin{equation}
M_{\nu }=%
\begin{pmatrix}
0_{3\times 3} & m_{\nu_D} & 0_{3\times 3} \\ 
m_{\nu_D}^{T} & 0_{3\times 3} & M \\ 
0_{3\times 3} & M^{T} & \mu%
\end{pmatrix}%
,
\end{equation}

and the submatrices are given by:
\begin{eqnarray}
m_{\nu_D} &=&\frac{y_{\nu }v_{\rho }v_{\Phi }}{\sqrt{2}\Lambda }\left( 
\begin{array}{ccc}
0 & r_1e^{i\theta} & r_1e^{i\theta} \\ 
-r_1e^{i\theta} & 0 & 1 \\ 
-r_1e^{i\theta} & -1 & 0%
\end{array}%
\right) ,\hspace{1cm}\hspace{1cm}M=y_{\chi }^{\left( L\right) }\frac{%
v_{\chi }}{\sqrt{2}}\left( 
\begin{array}{ccc}
1 & 0 & 0 \\ 
0 & 1 & 0 \\ 
0 & 0 & 1%
\end{array}%
\right) ,  \notag \\
\mu &=&\left( 
\begin{array}{ccc}
h_{1N}v_{\varphi } & h_{2N}v_{\xi }r_2 & h_{2N}v_{\xi} \\ 
h_{2N}v_{\xi }r_2 & h_{1N}v_{\varphi } & h_{2N}v_{\xi } \\ 
h_{2N}v_{\xi } & h_{2N}v_{\xi } & h_{1N}v_{\varphi }%
\end{array}%
\right) \frac{v_{\sigma }^{2}}{\Lambda ^{2}}.
\label{MR}
\end{eqnarray}

The light active masses arise from an inverse seesaw mechanism and the resulting physical neutrino mass matrices take the form~\cite{Mohapatra:1986bd,Malinsky:2005bi}:
\begin{eqnarray}
\widetilde{M}_{\nu} &=& m_{\nu_D}\left(M^T\right)^{-1}\mu M^{-1}m_{\nu_D}^T, \\
\widetilde{M}_{\nu}^{(1)} &=& \frac{1}{2}\mu-\frac{1}{2}\left(M+M^T\right), \\
\widetilde{M}_{\nu}^{(2)} &=& \frac{1}{2}\mu+\frac{1}{2}\left(M+M^T\right).
\end{eqnarray}

Here, $\widetilde{M}_{\nu}$ is the mass matrix for the active neutrino ($\nu_{\alpha}$), whereas $\widetilde{M}_{\nu}^{(1)}$ and $\widetilde{M}_{\nu}^{(2)}$ are the sterile neutrinos mass matrices.\\
Thus, the light active neutrino mass matrix is given by:
\begin{equation}
\widetilde{M}_{\nu} = \left(
\begin{array}{ccc}
 2 a^2 \left(y_2+z\right) & a \left(b \left(y_2+z\right)-a \left(y_1+y_2\right)\right) & -a \left(a \left(y_1+y_2\right)+b
   \left(y_2+z\right)\right) \\
 a \left(b \left(y_2+z\right)-a \left(y_1+y_2\right)\right) & z \left(a^2+b^2\right)-2 a b y_2 & a^2 z+a b \left(y_1-y_2\right)-b^2 y_2 \\
 -a \left(a \left(y_1+y_2\right)+b \left(y_2+z\right)\right) & a^2 z+a b \left(y_1-y_2\right)-b^2 y_2 & z \left(a^2+b^2\right)+2 a b y_1 \\
\end{array}
\right),
\label{ec:matrix-neutrino}
\end{equation}

where the effective parameters are defined as:
\begin{align}
a&=\frac{y_{\nu}v_{\rho}v_{\Phi}r_1}{y_{\chi}^{(L)}v_{\chi}\Lambda}e^{i\theta}, & b&= \frac{y_{\nu}v_{\rho}v_{\Phi}}{y_{\chi}^{(L)}v_{\chi}\Lambda}, & y_1&= \frac{v_{\sigma}^2h_{2N}v_{\xi}r_2}{\Lambda^2},\notag\\
y_2&= \frac{v_{\sigma}^2h_{2N}v_{\xi}}{\Lambda^2}, & z&= \frac{v_{\sigma}^2h_{1N}v_{\varphi}}{\Lambda^2}.\label{eq:efecpara}
\end{align}

From Eq.~\eqref{eq:efecpara}, we can observe that the effective parameters are not independent, obtaining the following relation,
\begin{equation}
a=b r_1e^{i\theta} \quad;\quad y_1=y_2r_2.
\end{equation}

%\blue{If we consider the Yukawa couplings as complex and the parameter $b$ as pure imaginary}, we obtain a cobimaximal texture for the mass matrix of the light-active neutrinos in the Eq. \eqref{ec:matrix-neutrino} and adapting the function $\chi^2$ from Eq. \eqref{ec:funtion_error} for the neutrino sector observables, we obtain the following error function:
In order to fit the effective neutrino sector parameters to successfully reproduce the experimental values of the neutrino mass squared splittings, the leptonic mixing angles and the leptonic Dirac CP phase, we proceed to minimize the following $\chi ^{2}$ function:
\begin{eqnarray}
\chi ^{2} &= & \frac{\left( \Delta m_{21}^{2\ \exp }-\Delta m_{21}^{2\ \text{th}}\right) ^{2}}{\sigma
_{\Delta m^2_{21}}^{2}}+\frac{\left( \Delta m_{31}^{2\ \exp }-\Delta m_{31}^{2\ \text{th}}\right) ^{2}}{\sigma
_{\Delta m^2_{31}}^{2}}+
\frac{\left( \sin^2\theta_{12}^{(l)\exp }-\sin^2\theta
_{12}^{(l)\text{th}}\right) ^{2}}{\sigma _{\sin^2 \theta^{(l)}_{12}}^{2}}\label{ec:error-neu}\\
\nn & & + \frac{\left( \sin^2\theta_{23}^{(l)\exp}-\sin^2\theta_{23}^{(l)\text{th}}\right) ^{2}}{\sigma_{\sin^2\theta^{(l)}_{23}}^{2}}
 + \frac{\left( \sin^2\theta_{13}^{(l)\exp }-\sin^2\theta_{13}^{(l)\text{th}}\right) ^{2}}{\sigma_{\sin^2\theta^{(l)}_{13}}^{2}}
+\frac{\left( \delta_{\text{CP}}^{\exp }-\delta_{\text{CP}}^{\text{th}}\right) ^{2}}{\sigma _{\delta_{\text{CP}}}^{2}}\;,  \notag
\end{eqnarray}

%which allows us to adjust the parameters of the model, 
where $\Delta m_{i1}^2$ (with $i= 2, 3)$ are the neutrino mass squared differences, $\sin\theta^{(l)}_{jk}$ is the sine function of the mixing angles (with $j,k=1,2,3$) and $\delta_{\text{CP}}$ is the CP violation phase. The supra indices represent the experimental (\enquote{exp}) and theoretical (\enquote{$\text{th}$}) values, and the $1\sigma$ are the experimental errors. However, as was done in the quark sector, the matrix \eqref{ec:matrix-neutrino} is not Hermitian.
Even so, due to its structure, the $\widetilde{M}_{\nu}\widetilde{M}_{\nu}^\dagger$ matrix is not simple to work with analytically, so numerical analysis is also carried out in the neutrino sector. Therefore, after minimizing Eq.~\eqref{ec:error-neu} using the matrix $\widetilde{M}_{\nu}\widetilde{M}_{\nu}^\dagger$, we get the following values for the model parameters:
\begin{align}
a&= (0.521\pm 0.022)e^{-3.027i}, & b&= (1.080\pm 0.041)i, & y_1&=-0.086\pm 0.007\ \text{eV},\notag\\
y_2&= 2.05\pm 0.12\ \text{eV}, & z&= -2.30\pm 0.14\ \text{eV}.\label{ec:bestpara}
\end{align}

The diagonalization of the matrix \eqref{ec:matrix-neutrino} gives us a determinant equal to zero, so we obtain the following eigenvalues:
\begin{eqnarray}
 m_1^2 &=& 0 ,\\
m_2^2 &=& \bigg| \left(a^2\left(y_2+2 z\right)+a b \left(y_1-y_2\right)+b^2 z\right. \notag\\
& & -\left.\left.\sqrt{2 a^2 y_1^2 \left(a^2+b^2\right)+2 a y_2 y_1 (a-b) \left(2 a^2+a b+b^2\right)+y_2^2 \left(a^2+b^2\right) \left(3 a^2+2 a b+b^2\right)}\right)\right\lvert^2, \\
m_3^2 &=&\bigg| \left(a^2\left(y_2+2 z\right)+a b \left(y_1-y_2\right)+b^2 z\right. \notag \\
&  & +\left.\sqrt{2 a^2 y_1^2 \left(a^2+b^2\right)+2 a y_2 y_1 (a-b) \left(2 a^2+a b+b^2\right)+y_2^2 \left(a^2+b^2\right) \left(3 a^2+2 a b+b^2\right)}\right) \bigg|^2.   
\end{eqnarray}

With the values of the parameters of our best-fit point of the Eq. \eqref{ec:bestpara}, we obtain the results of the neutrino sector that are shown in the Table \ref{table:neutrinos_value}, together with the experimental values in the range of $1\sigma$ and $3\sigma$, whose experimental data were taken from \cite{deSalas:2020pgw}. In the Table \ref{table:neutrinos_value}, we can see that the difference of the square of the neutrino masses ($\Delta m_{21}^2$, $\Delta m_{31}^2$) and the of solar and reactor mixing neutrinos ($\sin^2\theta_{12}^{(l)}$, $\sin^2\theta_{13}^{(l)}$) are in the range of $1\sigma$, while the angle of the atmospheric neutrino ($\sin^2\theta_{23}^{(l)}$) is in the range of $2\sigma$ and the CP violation phase ($\delta_{\text{CP}}$) is in the range of $3\sigma$.

\begin{table}[tp]
%\resizebox{13cm}{!}{
\begin{tabular}{c|c|cccccc}
\toprule[0.13em] Observable & range & $\Delta m_{21}^{2}$ [$10^{-5}$eV$^{2}$]
& $\Delta m_{31}^{2}$ [$10^{-3}$eV$^{2}$] & $\sin^2\theta^{(l)}_{12}/10^{-1}$
& $\sin^2\theta^{(l)}_{13}/10^{-3}$ & $\sin^2\theta^{(l)}_{23}/10^{-1}$ & $%
\delta^{(l)}_{\text{CP}} (^{\circ })$ \\ \hline
Experimental & $1\sigma$ & $7.50_{-0.20}^{+0.22}$ & $2.55_{-0.03}^{+0.02}$ & 
$3.18\pm 0.16$ & $2.200_{-0.062}^{+0.069}$ & $5.74\pm 0.14$ & $%
194_{-22}^{+24}$ \\ 
Value & $3\sigma$ & $6.94-8.14$ & $2.47-2.63 $ & $2.71-3.69$ & $2.000-2.405$
& $4.34-6.10$ & $128-359$ \\ \hline
Fit & $1\sigma-3\sigma$ & $7.498$ & $2.54$ & $3.02$ & $2.191$ & $5.93$ & $246.4$\\ \hline
%\bottomrule[0.13em] &  &  &  &  &  &  & 
\end{tabular}
\caption{Model predictions for the scenario of normal order (NO) neutrino
mass. The experimental values are taken from Ref. \protect\cite{deSalas:2020pgw}}
\label{table:neutrinos_value}
\end{table}

\begin{figure}[]
\centering
\subfigure[]{
\includegraphics[scale=0.35]{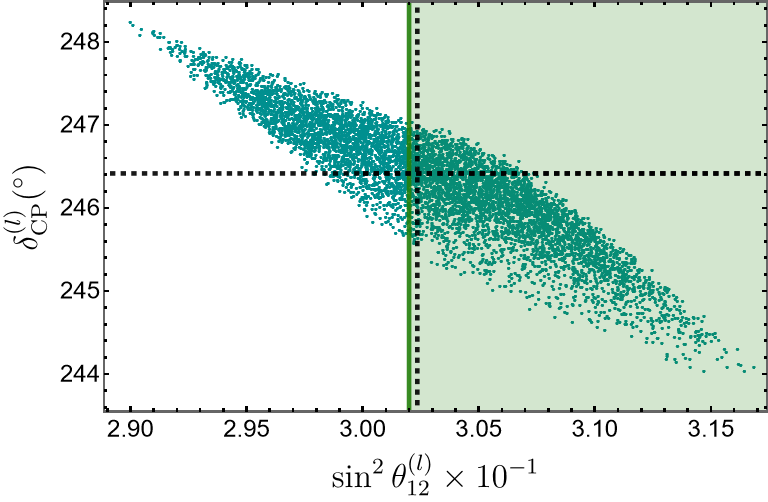}
} \quad
\subfigure[]{
\includegraphics[scale=0.35]{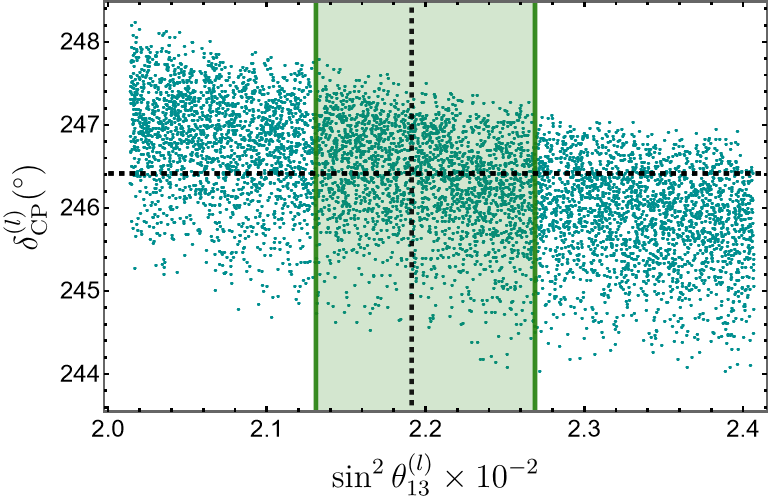}
}\quad
\subfigure[]{
\includegraphics[scale=0.35]{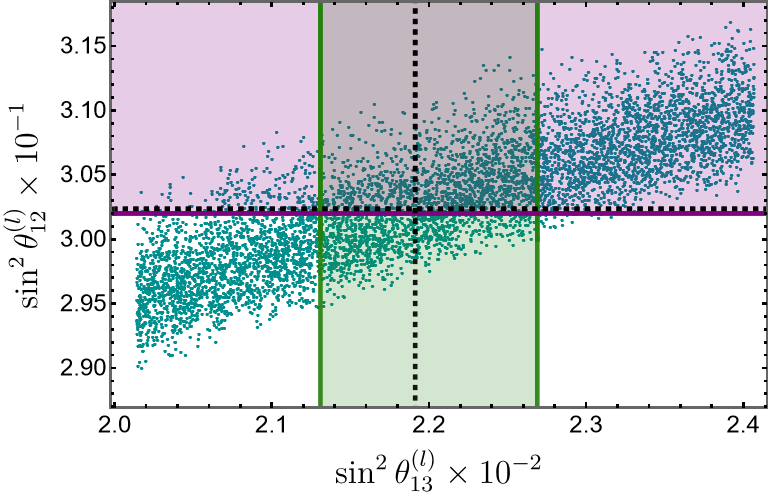}
}
\subfigure[]{
\includegraphics[scale=0.35]{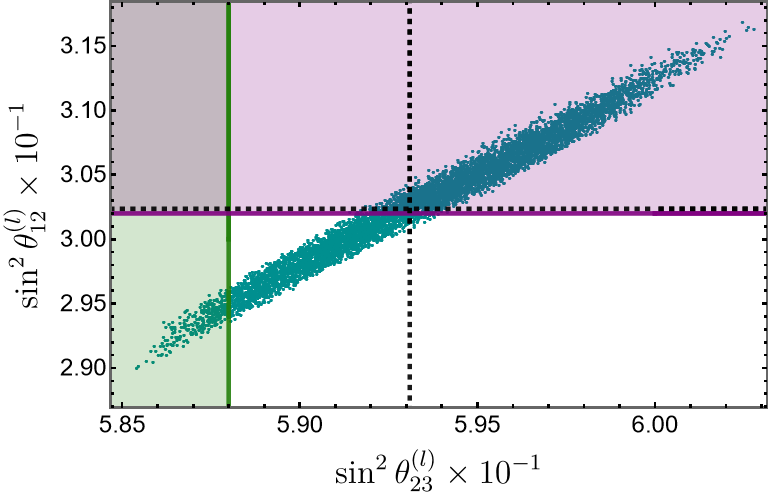}
}
\subfigure[]{
\includegraphics[scale=0.35]{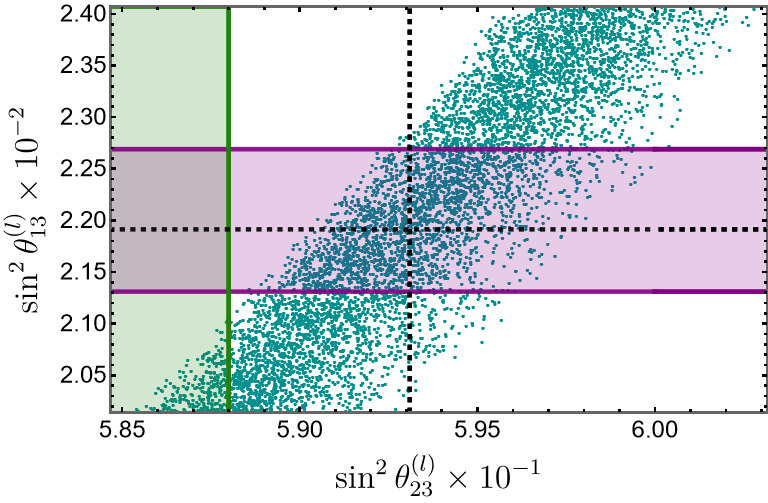}
}
\subfigure[]{
\includegraphics[scale=0.35]{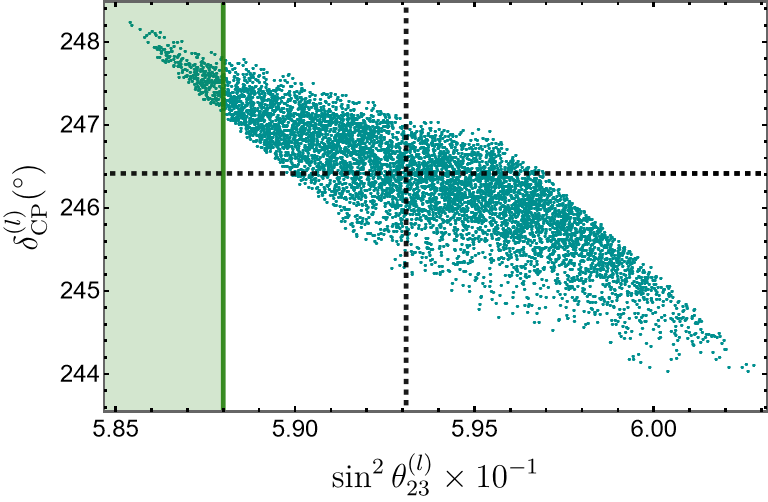}
}
\caption{Correlation between the mixing angles of the neutrino sector and the CP violation phase obtained with our model. The green and purple bands represent the $1\sigma$ range in the experimental values, while the dotted line (black) represents the best-fit point of our model.}
\label{fig:neutrino-corr}
\end{figure}

Fig. \ref{fig:neutrino-corr} shows the correlation between the leptonic Dirac CP violating phase and the neutrino mixing angles as well as the correlations among the leptonic mixing angles, where the green and purple background fringes represent the $1\sigma$ range of the experimental values and the black bands the dotted lines represent our best-fit point for each observable. In Fig. \ref{fig:neutrino-corr}, we see that for the mixing angles, we can get values in the $1\sigma$ range, while for the CP violating phase, we obtain values up to $3\sigma$, where each lepton sector observable is obtained in the following range of values: $0.290\leq \sin^2\theta^{(l)}_{12}\leq 0.317$, $0.0201\leq \sin^2\theta^{(l)}_{13}\leq 0.0241$, $0.584\leq \sin^2\theta^{(l)}_{23}\leq 0.603$ and $244^\circ\leq \delta_{\text{CP}}\leq 248^\circ$.

%%%%%%%%%%%%%%%%%%%%%%%%%%%%%%%%%%%%%%%%%%%%%%%%%%%%%%%%%%%%%%%%%%%%%%%%%%%%%%%%%%%%%%%%

\section{\label{scalarsector}Scalar potential for the $SU(3)_{L}$ triplets.}\label{scalar}

In this section we will discuss the low energy scalar potential as well as the resulting scalar mass spectrum. As previously mentioned, the discrete groups are spontaneously broken at energy scale much larger than the scale of spontaneoys breaking of the $SU(3)_C\times SU(3)_L\times U(1)_X\times U(1)_{L_g}$ symmetry, then implying that the mixing angles between the gauge singlet scalar fields and the $SU(3)_L$ scalar triplets are strongly suppressed by the ratios between their VEVs (in the scenario of quartic scalar couplings of the same order of magnitude), as follows from the method of method of recursive expansion proposed in \cite{Grimus:2000vj}. Because of this reason, the singlet scalar fields do not have a relevant impact in the electroweak precision observables since they do not couple with the SM gauge bosons and their mixing angles with the neutral components of the $SU(3)_L$ scalar triplets are very small. Therefore, under the aforementioned considerations, we restrict of our analysis of the scalar spectrum to the one resulting from the scalar potential built from $SU(3)_L$ triplets of our model, which are the relevant scalar degrees of freedom at energies corresponding to the scale were the $SU(3)_C\times SU(3)_L\times U(1)_X\times U(1)_{L_g}$ symmetry is spontaneously broken. It is worth mentioning that the most general renormalizable potential invariant under $S_4$ which we can write with the four triplets of the Eqs. (\ref{triplet}) is given by
\begin{eqnarray}\label{Vhiggs}
V &=&-\mu _{\chi }^{2}(\chi ^{\dagger }\chi ) + \mu
_{\eta _{1}}^{2}\eta _{1}^{\dagger }\eta _{1}-\mu _{\eta _{2}}^{2}\eta
_{2}^{\dagger }\eta _{2} -\mu _{\rho }^{2}(\rho
^{\dagger }\rho )+ \varkappa~ \eta _{2}\rho \chi +\lambda_{1}(\chi ^{\dagger }\chi )(\chi ^{\dagger }\chi )+\lambda_{2}(\eta ^{\dagger }\eta )_{\mathbf{1}}(\eta ^{\dagger }\eta )_{\mathbf{1}}
\notag \\
&&+\lambda _{3}(\eta ^{\dagger }\eta )_{\mathbf{1}^{\prime }}(\eta ^{\dagger
}\eta )_{\mathbf{1}^{\prime }}+\lambda _{4}(\eta ^{\dagger }\eta )_{\mathbf{2%
}}(\eta ^{\dagger }\eta )_{\mathbf{2}} +\lambda _{5}(\eta ^{\dagger }\eta )_{\mathbf{1}}(\chi ^{\dagger }\chi)+\lambda _{6}\left[ (\eta ^{\dagger }\chi )(\chi ^{\dagger }\eta )\right] _{%
\mathbf{1}}\\
&&+\lambda _{7}(\rho ^{\dagger }\rho )^{2}+\lambda _{8}(\eta
^{\dagger }\eta )_{\mathbf{1}}(\rho ^{\dagger }\rho )+\lambda _{9}(\chi
^{\dagger }\chi )(\rho ^{\dagger }\rho )+h.c,  \notag
\end{eqnarray}
where $\eta_{1}$ is an inert $SU(3)_L$ scalar triplet, the $\mu$’s are mass parameters and $\varkappa$ is the trilinear scalar coupling, has dimensions of mass, while $\lambda$’s are the quartic dimensionless couplings. Furthermore, the  minimization conditions of the scalar potential yield the following relations

\begin{eqnarray}
\nonumber \mu_{\chi}^2 &=&-\frac{\varkappa}{\sqrt{2}} \frac{v_{\eta _2} v_{\rho }}{v_{\chi}}
+\frac{\lambda_5}{2}v_{\eta _2}^2 
+\frac{\lambda_9}{2}v_{\rho }^2 
+\lambda_1 v_{\chi}^2,\\
\label{constraints} \mu_{\eta_2}^2 &=& -
\frac{\varkappa}{\sqrt{2}} \frac{v_{\rho } v_{\chi }}{v_{\eta _2}}+\frac{\lambda_8}{2} v_{\rho}^2 +\frac{\lambda_5}{2} v_{\chi}^2 +\left(\lambda_2+\lambda_4\right)  v_{\eta _2}^2, \\
\nonumber \mu_{\rho}^2 &=& -
\frac{\varkappa}{\sqrt{2}} \frac{v_{\eta _2} v_{\chi }}{v_{\rho }}+\frac{\lambda_8}{2}v_{\eta_2}^2+\frac{\lambda_9}{2}v_{\chi }^2+\lambda_7 v_{\rho}^2.
\end{eqnarray}

\begin{figure}
\subfigure[]{\includegraphics[scale=0.3]{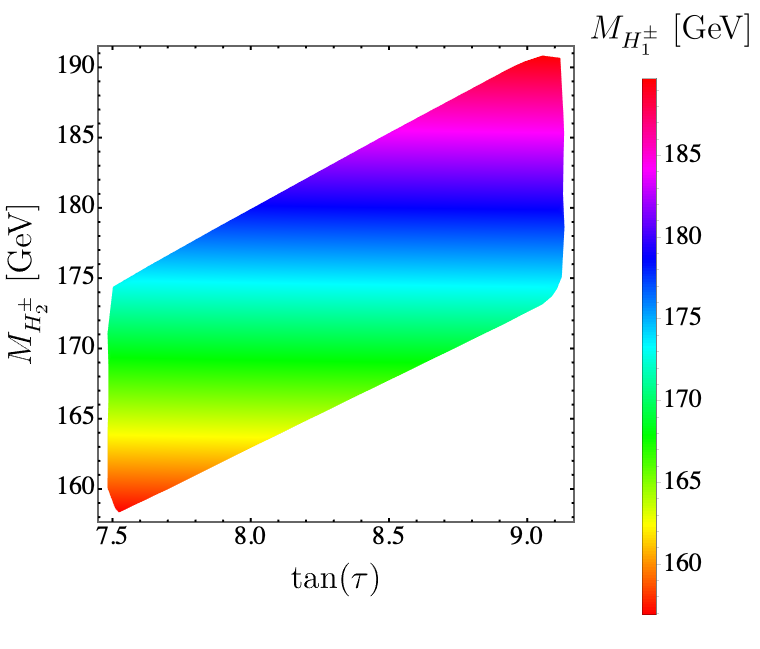}}
%\subfigure[]{\includegraphics[scale=0.3]{graf/deltaMHC2MHC1}}
%\subfigure[]{\includegraphics[scale=0.3]{graf/deltaMHC2MA01}}
%\subfigure[]{\includegraphics[scale=0.3]{graf/deltaMHC1MA01.png}}
\subfigure[]{\includegraphics[scale=0.3]{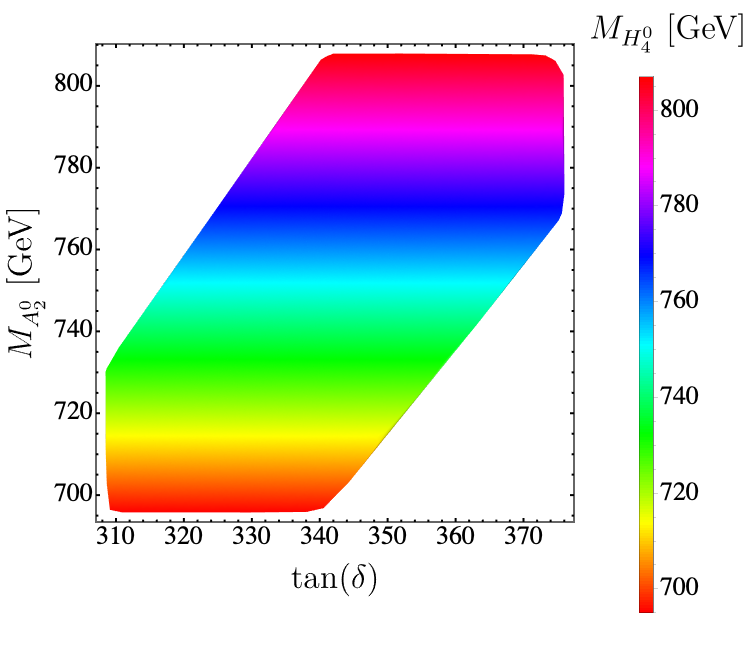}}
%\subfigure[]{\includegraphics[scale=0.3]{graf/betaMH04MA02}}
% \subfigure[]{\includegraphics[scale=0.3]{graf/alfaMA02MH04}}
\caption{Correlations between mixing angles and the masses of the physical charged scalar, neutral scalar/pseudoscalar fields.}
\label{mangles}
\end{figure}

After spontaneous symmetry breaking, the Higgs mass spectrum comes from the diagonalization of the squared mass matrices (see appendix \ref{apex:matrixscalar}). The mixing angles\footnote{The symbol $\beta$ is used in the scalar potential for the $SU(3)_{L}$ triplets as one of the mixing angles, it is a symbol different from the definition of the operator electric charge.} for the physical eigenstates are:
\begin{eqnarray} \label{mixingangles}
\tan\alpha &=& \frac{v_{\eta_2}}{v_{\chi}}, \quad \tan\beta \ =\ \frac{v_{\rho}}{v_{\chi}}, \quad \tan\tau \ =\ \frac{v_{\eta_2}}{v_{\rho}}, \quad \tan\delta = \frac{v_{\chi}}{v_{\rho} \sin\tau}, \quad \tan2\vartheta = \frac{2\sqrt{2} \lambda_8}{\varkappa v_\chi}\frac{v_{\eta_2}^2 v_\rho^2}{v_\rho^2-v_{\eta_2}^2 },
\end{eqnarray}  
The Fig.~\ref{mangles}, presents correlation plots demonstrating the relationships between mixing angles and physical scalar masses, these correlation plots have been obtained by varying the point of best fit of the potential sector parameters around $10\%$. These plots highlight specific correlations, such as the correlation between the $\tau$ angle and charged fields, and the correlation between the $\delta$ angle and charged and pseudo-scalar fields. These correlations provide invaluable insights into the interactions among scalar fields and enhance our understanding of particle properties and the relationships between their masses and mixing angles. Moreover, similar correlations were observed for other mixing angles. The correlation analyses of the mixing angles offer valuable information regarding the underlying theoretical structure and relationships within the model.

We find that the charged sector is composed of two Goldstone bosons and three massive charged scalars.
\begin{eqnarray}
M_{G_{1}^{\pm }}^{2} &=& M_{G_{2}^{\pm }}^{2}=0, \\
M_{H_{1}^{\pm }}^{2} &=&\frac{\varkappa v_{\eta _{2}}v_{\rho }}{\sqrt{2}v_{\chi }}%
+\frac{\varkappa v_{\eta _{2}}v_{\chi }}{\sqrt{2}v_{\rho }} ,\\
M_{H_{2}^{\pm }}^{2} &=&\frac{\varkappa v_{\eta _{2}}v_{\chi }}{\sqrt{2}v_{\rho }}%
+\frac{\varkappa v_{\rho }v_{\chi }}{\sqrt{2}v_{\eta _{2}}}, \\
M_{H_{3}^{\pm }}^{2} &=&\mu_{\eta_1}^{2}+\left( \lambda _{2}-\lambda _{4}\right)
v_{\eta _{2}}^{2}+\frac{\lambda _{8}v_{\rho }^{2}}{2}+\frac{\lambda
_{5}v_{\chi }^{2}}{2}.
\end{eqnarray}

\begin{comment}
\begin{eqnarray}
M^2_{H_1^{\pm}} &=&  \frac{\varkappa}{\sqrt{2}}a v_{\eta_2}  \csc (\beta )
   \sec (\beta ),\\
M^2_{H_2^{\pm}} &=&\frac{\varkappa}{\sqrt{2}} v_\chi \csc (\tau ) \sec (\tau ),\\
 M^2_{H_3^{\pm}}  &= &  \mu_{\eta_1}^{2}+ v_\chi^2 \left(
 \frac{\lambda_5}{2} + \left(\lambda_2-\lambda_4 \right) \tan ^2(\alpha )
 +\frac{\lambda_8}{2} \tan^2(\beta)\right).
\end{eqnarray}
\end{comment}

The Goldstone bosons come only from mixing between $\rho_3^{\pm}$ and $\eta_{22}^\pm$, through the angle $\tau$, while $\chi_2^\pm$ is a massive field charged, the other two bulk fields correspond to the blending of $\rho_1^\pm$ and the charged component of the scalar inert $\eta_{21}^\pm$ by blending angle $\beta$, i.e,
\begin{equation}
\begin{array}{lll}
G_{1}^{\pm }=\cos \tau \ \rho _{3}^{\pm }-\sin \tau \ \eta _{22}^{\pm }, & 
& G_{2}^{\pm }=\sin \tau \ \rho _{3}^{\pm }+\cos \tau \ \eta _{22}^{\pm },
\\ 
H_{1}^{\pm }=\cos \beta \ \rho _{1}^{\pm }-\sin \beta \ \eta _{21}^{\pm }, & 
& H_{2}^{\pm }=\sin \beta \ \rho _{1}^{\pm }+\cos \beta \ \eta _{21}^{\pm },
\\ 
H_{3}^{\pm }=\chi _{2}^{\pm }. &  & 
\end{array}
\end{equation}

The physical mass eigenvalues of the CP odd scalars $A_ 1^0, \ A_2^0$ and the Goldstone bosons $G_1^0, \ G_2^0$ can be written as:
\begin{eqnarray}
M_{G_{1}^{0}}^{2} &=&M_{G_{2}^{0}}^{2} \ = \ 0, \\
M_{A_{1}^{0}}^{2} &=&\frac{\varkappa v_{\eta _{2}}v_{\rho }}{\sqrt{2}v_{\chi }}+%
\frac{\varkappa v_{\eta_2} v_{\chi}}{\sqrt{2} v_{\rho}} +\frac{\varkappa v_{\rho} v_{\chi}}{\sqrt{2} v_{\eta_2}}, \\
M_{A_{2}^{0}}^{2} &=&\mu _{\eta_1}^{2}+v_{\eta _{2}}^{2}\left( \lambda_{2}-2\lambda _{3}-\lambda_{4}\right) +\frac{\lambda_{8}v_{\rho }^{2}}{2}+%
\frac{\lambda _{5}v_{\chi }^{2}}{2}.
\end{eqnarray}

\begin{comment}
\begin{eqnarray}
M_{A_{1}^{0}}^{2} &=& \frac{\varkappa}{\sqrt{2}} \csc (\tau ) \bigg(
v_{\eta_2} \cot (\delta )+v_\chi \sec (\tau )\bigg), \\
M_{A_{2}^{0}}^{2} &=&\mu _{\eta_1}^{2}+v_\chi^2 \left(
 \frac{\lambda_5}{2} + \left(\lambda_2-2\lambda_3-\lambda_4 \right) \tan ^2(\alpha )
 +\frac{\lambda_8}{2} \tan^2(\beta)\right).
\end{eqnarray} 
\end{comment}

We have the following relationship between the original physical eigenstates:
\begin{equation}
\begin{array}{lll}
G_{1}^{0}=\cos \tau \ \zeta _{\rho }-\cos \delta \cos \tau \ \zeta _{\eta_1}-\sin \tau \ \zeta _{\eta _{2}}, &  & G_{2}^{0}=\cos \delta \ \zeta _{\rho
}+\sin \delta \ \zeta _{\eta _{1}}, \\ 
A_{1}^{0}=\sin \tau \ \zeta _{\rho }-\cos \delta \sin \tau \ \zeta _{\eta
_{1}}+\cos \tau \ \zeta _{\eta _{2}}, &  & A_{2}^{0}=\zeta _{\chi },%
\end{array}
\end{equation}
where consider the following limit $v_\chi\gg v_{\rho},v_{\eta_2}$.

The masses of the light and heavy eigenstates for CP even scalars are given as:
\begin{eqnarray}
M^2_h  &=& \varkappa \frac{\left( v_{\eta_2}^2 + v_\rho^2\right) v_\chi}{2\sqrt{2}v_{\eta_2} v_\rho } - \frac{1}{2\sqrt{2}v_{\eta_2} v_\rho }  \sqrt{\varkappa^2\left( v_{\eta_2}^2 - v_\rho^2\right)^2 v_\chi^2 +8 \lambda_8^2 v_{\eta_2}^4  v_\rho^4} ,\\
M_{H_1^0}^2 &=&  \varkappa \frac{\left( v_{\eta_2}^2 + v_\rho^2\right) v_\chi}{2\sqrt{2}v_{\eta_2} v_\rho } +\frac{1}{2\sqrt{2}v_{\eta_2} v_\rho }  \sqrt{\varkappa^2\left( v_{\eta_2}^2 - v_\rho^2\right)^2 v_\chi^2 +8 \lambda_8^2v_{\eta_2}^4  v_\rho^4} ,\\
M_{H_2^0}^2&=& \mu _{\eta _1}^2+\left(\lambda _2+\lambda _4\right)
   v_{\eta _2}^2+\frac{\lambda_8}{2}v_{\rho}^2+\frac{\lambda _5}{2} v_{\chi}^2,\\
M_{H_3^0}^2 &=&  2 \lambda _1 v_{\chi }^2.
\end{eqnarray}

\begin{comment}
\begin{eqnarray}
M^2_h  &=& \frac{\varkappa }{2\sqrt{2}} v_\chi\csc (\tau ) \sec (\tau )
   \bigg( 1+\sec (2 \vartheta ) \cos (2
   \tau )\bigg) ,\\
M_{H_1^0}^2 &=&  \frac{\varkappa }{2\sqrt{2}} v_\chi \csc (\tau ) \sec (\tau )
   \bigg( 1-\sec (2 \vartheta ) \cos (2
   \tau )\bigg) ,\\
M_{H_2^0}^2&=&  \mu_{\eta_1}^{2}+ v_\chi^2 \left(
 \frac{\lambda_5}{2} + \left(\lambda_2+\lambda_4 \right) \tan ^2(\alpha )
 +\frac{\lambda_8}{2} \tan^2(\beta)\right),\\
M_{H_3^0}^2 &=&  2 \lambda_1 v_{\chi }^2.
\end{eqnarray}  
\end{comment}

The lighter mass eigenstate $h$ is identified as the SM Higgs boson. The two mass eigenstates $h$ and $H_1^0$ are related with the $\xi_{\eta_2}$ and $\xi_\rho$ fields through the rotation angle $\vartheta$ as:
 \begin{eqnarray}
h&\simeq &  \xi_{\eta_2}\cos\vartheta -  \xi_{\rho}  \sin\vartheta ,\\
 H_1^0&\simeq & \xi_{\eta_2}\sin\vartheta +  \xi_{\rho}  \cos\vartheta, 
\end{eqnarray}
while the heavier fields are related as $H_2^0 \simeq \xi_{\eta_1}$ and $H_3^0 \simeq \xi_\chi$.

Finally, for the pseudoscalar and scalar neutral complex fields, we have composed the mixture of the Imaginary and Real parts of $\eta_{31}^0,\ \eta_{32}^0,\ \chi_{1}^0$, respectively,
\begin{eqnarray}
M_{G_{3}^{0}}^{2} &=&M_{G_{4}^{0}}^{2}\ = \ 0 \\
M_{A_{3}^{0}}^{2} &=& \sqrt{2}\varkappa \left(
\frac{v_{\eta _{2}}v_{\rho }}{v_{\chi }}+
\frac{v_{\rho }v_{\chi }}{v_{\eta _{2}}}
\right)
-\lambda _{6}\left(
v_{\chi }^{2}+v_{\eta _{2}}^{2}\right),  \\
M_{A_{4}^{0}}^{2} &=&2\mu _{\eta_1}^{2}
+2\left(\lambda_2-\lambda_4 \right)v_{\eta _{2}}^{2}
+\lambda_8 v_\rho^2 + \left(\lambda_5-\lambda_6 \right) v^2_\chi,\\
M_{H_{4}^{0}}^{2} &=& \sqrt{2}\varkappa\left(
\frac{v_{\eta _{2}}v_{\rho }}{v_{\chi }}+
\frac{v_{\rho }v_{\chi }}{v_{\eta _{2}}}
\right)
+\lambda _{6}\left(
v_{\chi }^{2}+v_{\eta _{2}}^{2}\right) , \\
M_{H_{5}^{0}}^{2} &=&2\mu _{\eta_1}^{2}+2\left(\lambda _{2}-\lambda_{4}\right) v_{\eta _{2}}^{2}+\lambda _{8}v_{\rho }^{2}+\left(\lambda _{5}+\lambda _{6}\right) v_{\chi }^{2}.
\end{eqnarray}

\begin{comment}
\begin{eqnarray}
M_{A_{3}^{0}}^{2} &=& v_{\eta_2} \csc^2(\alpha) \bigg(
\sqrt{2} \varkappa \tan(\alpha) -  \lambda_6 v_{\eta_2}
\bigg),\\
M_{A_{4}^{0}}^{2} &=&2\mu _{\eta_1}^{2} + v^2_\chi \bigg(\lambda_5 - \lambda_6 + 2\left(\lambda_2-\lambda_4  \right) \tan^2(\alpha) + \lambda_8 \tan^2(\beta)\bigg),\\
M_{H_{4}^{0}}^{2} &=& v_{\eta_2} \csc^2(\alpha) \bigg(
\sqrt{2} \varkappa \tan(\alpha) +  \lambda_6 v_{\eta_2}
\bigg),\\
M_{H_{5}^{0}}^{2} &=&2\mu _{\eta_1}^{2} + v^2_\chi \bigg(\lambda_5 + \lambda_6 + 2\left(\lambda_2-\lambda_4  \right) \tan^2(\alpha) + \lambda_8 \tan^2(\beta)\bigg).
\end{eqnarray} 
\end{comment}

In the physical eigenstates, there are two Goldstone bosons and one pseudoscalar massive boson, and a scalar from the mixture of the complex neutral part of $\eta_1$ and $\eta_2$, while,
\begin{equation}
\begin{array}{lll}
G_{3}^{0}=\sin \alpha\ \text{Im}\eta_{31}^0-\cos \alpha \ \text{Im}\eta _{32}^{0}, &  & G_{4}^{0}=-\sin \alpha \ \text{Re}\eta_{31}^{0}+\cos \alpha \ \text{Re}\eta_{32}^{0}, \\ 
A_{3}^{0}=\cos \alpha \ \text{Im}\eta _{31}^{0}+\sin \alpha \ \text{Im}\eta_{32}^{0}, &  & H_{4}^{0}=\cos \alpha \ \text{Re}\eta _{31}^{0}+\sin\alpha \ \text{Re}\eta_{32}^{0}, \\ 
A_{4}^{0}=\text{Im}\chi_{1}^{0}, &  & H_{5}^{0}=\text{Re}\chi_{1}^{0}.
\end{array}
\end{equation}

\begin{figure}
\subfigure[]{\includegraphics[scale=0.4]{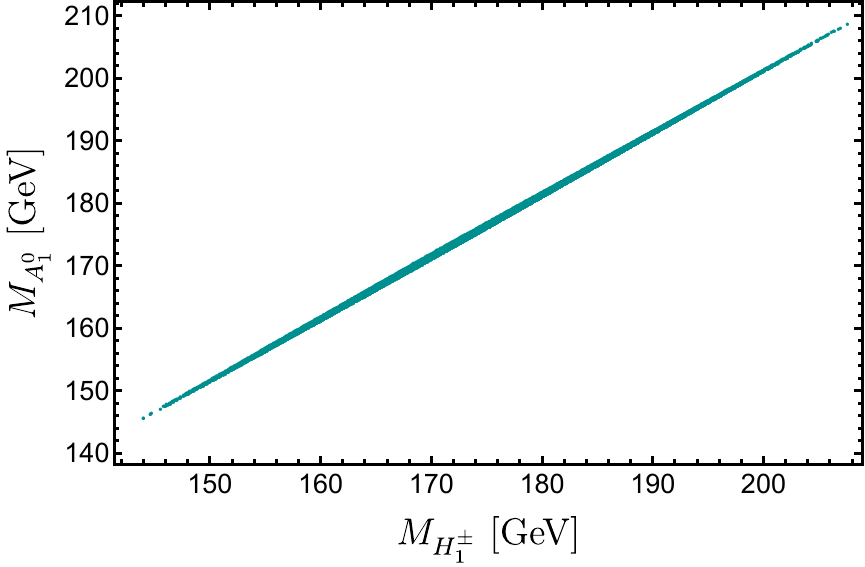}}
\subfigure[]{\includegraphics[scale=0.4]{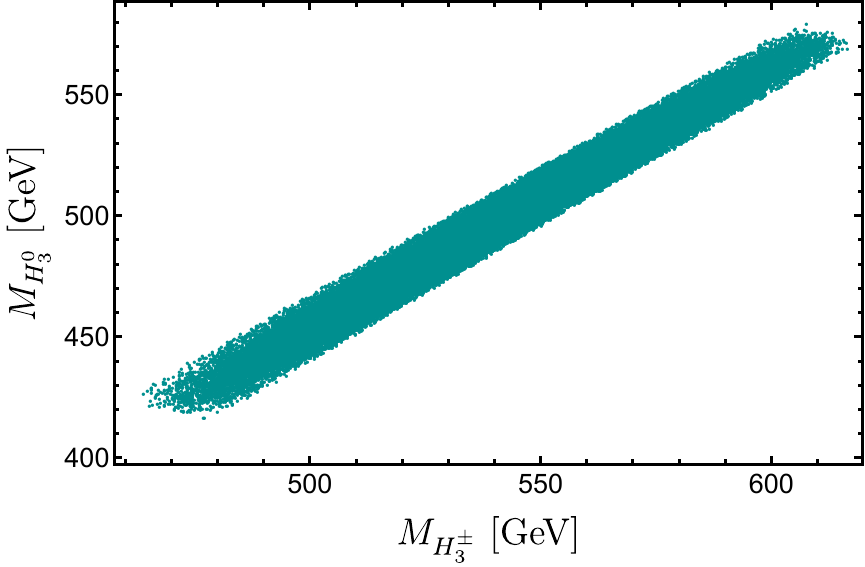}}
\caption{Correlation plot between the pseudoscalar neutral, scalar neutral, and scalar charged masses.}
\label{fig:scalar-sector}
\end{figure}

In our model, the physical scalar masses can also be expressed %for the physical scalar mass spectrum, they can be expressed 
%as a function of 
in terms of the scalar mixing angles, as shown in table \ref{tab:massanglesmix}, where the light scalar field $h$, is identified as the SM-like Higgs boson, additionally six charged fields $\left(H_1^{\pm}, H_2^{\pm},H_3^{\pm} \right)$, five CP even $\left(H_1^0,H_2^0,H_3^0,H_4^0, H_5^0 \right)$ and four CP odd fields $\left(A_1^0,A_2^0,A_3^0, A_4^0 \right)$. Fig.~\ref{fig:scalar-sector}, displays %corresponds to cases of 
the correlation between the charged and neutral scalar masses. Fig.~\ref{fig:scalar-sector} \textcolor{red}{(a)}, shows a linear correlation between of the masses between the charged field and the pseudoscalar neutral field, $H_1^{\pm}$ and $A_1^0$ respectively, in Fig.~\ref{fig:scalar-sector} \textcolor{red}{(b)}, a linear correlation between the masses of the pseudoscalar and scalar neutral field, $A_3^0$ and $H_3^0$ respectively. The charged Goldstone bosons $\left(G_1^{\pm},G_2^{\pm} \right)$ are related to the longitudinal components of the $W^{\pm}$ and $W'^{\pm}$ gauge bosons respectively; while the neutral Goldstone bosons $\left(G_1^0,G_2^0,G_3^0,G_4^0\right)$ are associated to the longitudinal components of the $Z$, $Z'$, $K^0$ and $K'^0$ gauge bosons.

\begin{table}[h]
    \centering
    \begin{tabular}{|c|c|}
        \hline
        \text{Scalar} & \text{Masses} \\
        \hline
        $M_{G_{1}^{\pm }}^{2}, \, M_{G_{2}^{\pm }}^{2}$ & $0$ \\
        $M_{H_1^{\pm}}^{2}$ & $\frac{\varkappa}{\sqrt{2}}a v_{\eta_2}  \csc (\beta ) \sec (\beta )$ \\
        $M_{H_2^{\pm}}^{2}$ & $\frac{\varkappa}{\sqrt{2}} v_\chi \csc (\tau ) \sec (\tau )$ \\
        $M_{H_3^{\pm}}^{2}$ & $\mu_{\eta_1}^{2} + v_\chi^2 \left( \frac{\lambda_5}{2} + \left(\lambda_2 - \lambda_4 \right) \tan^2(\alpha ) + \frac{\lambda_8}{2} \tan^2(\beta) \right)$ \\
        \hline
        $M_{G_{1}^{0}}^{2}, \, M_{G_{2}^{0}}^{2}$ & $0$ \\
        $M_{A_{1}^{0}}^{2}$ & $\frac{\varkappa}{\sqrt{2}} \csc (\tau ) \left( v_{\eta_2} \cot (\delta ) + v_\chi \sec (\tau ) \right)$ \\
        $M_{A_{2}^{0}}^{2}$ & $\mu_{\eta_1}^{2} + v_\chi^2 \left( \frac{\lambda_5}{2} + \left( \lambda_2 - 2\lambda_3 - \lambda_4 \right) \tan^2(\alpha ) + \frac{\lambda_8}{2} \tan^2(\beta) \right)$ \\
        \hline
        $M_{h}^{2}$ & $\frac{\varkappa }{2\sqrt{2}} v_\chi \csc (\tau ) \sec (\tau ) \left( 1 + \sec (2 \vartheta ) \cos (2 \tau ) \right)$ \\
        $M_{H_1^0}^{2}$ & $\frac{\varkappa }{2\sqrt{2}} v_\chi \csc (\tau ) \sec (\tau ) \left( 1 - \sec (2 \vartheta ) \cos (2 \tau ) \right)$ \\
        $M_{H_2^0}^{2}$ & $\mu_{\eta_1}^{2} + v_\chi^2 \left( \frac{\lambda_5}{2} + \left( \lambda_2 + \lambda_4 \right) \tan^2(\alpha ) + \frac{\lambda_8}{2} \tan^2(\beta) \right)$ \\
        $M_{H_3^0}^{2}$ & $2 \lambda_1 v_{\chi }^2$ \\
        \hline
        $M_{G_{3}^{0}}^{2}, \, M_{G_{4}^{0}}^{2}$ & $0$ \\
        $M_{A_{3}^{0}}^{2}$ & $v_{\eta_2} \csc^2(\alpha) \left( \sqrt{2} \varkappa \tan(\alpha) - \lambda_6 v_{\eta_2} \right)$ \\
        $M_{A_{4}^{0}}^{2}$ & $2 \mu_{\eta_1}^{2} + v_\chi^2 \left( \lambda_5 - \lambda_6 + 2 \left( \lambda_2 - \lambda_4 \right) \tan^2(\alpha) + \lambda_8 \tan^2(\beta) \right)$ \\
        $M_{H_{4}^{0}}^{2}$ & $v_{\eta_2} \csc^2(\alpha) \left( \sqrt{2} \varkappa \tan(\alpha) + \lambda_6 v_{\eta_2} \right)$ \\
        $M_{H_{5}^{0}}^{2}$ & $2 \mu_{\eta_1}^{2} + v_\chi^2 \left( \lambda_5 + \lambda_6 + 2 \left( \lambda_2 - \lambda_4 \right) \tan^2(\alpha) + \lambda_8 \tan^2(\beta) \right)$ \\
        \hline
    \end{tabular}
    \caption{Physical mass spectrum of the scalars in terms of mixing angles.}
    \label{tab:massanglesmix}
\end{table}

\begin{figure}
\includegraphics[scale=0.35]{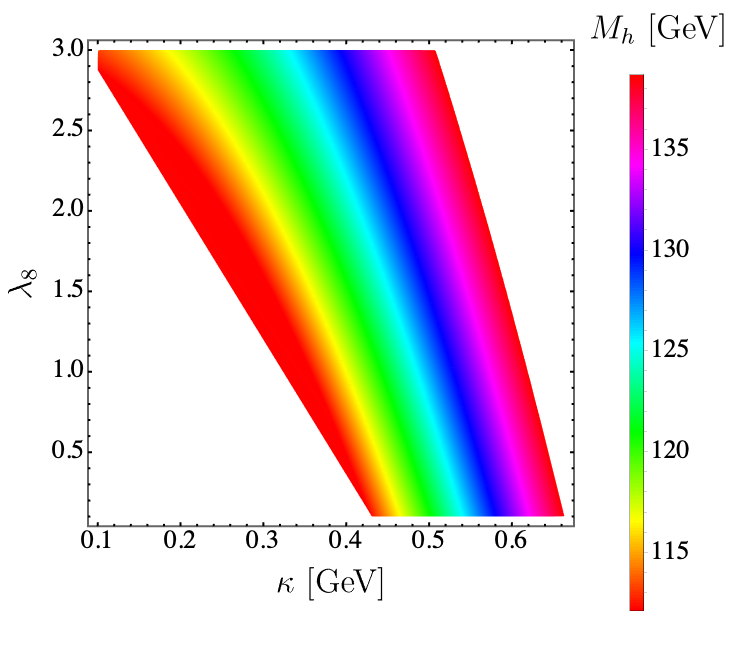}
\caption{Correlation color plot between the trilinear scalar parameter $\varkappa$ and the quartic scalar coupling $\lambda_8$ consistent with the SM-like Higgs mass.}
\label{fig:kappavslambda8}
\end{figure}

In analyzing the scalar sector of the model, %setting up our model, 
it should be noted that in the potential of Eq.~\eqref{Vhiggs}, the quartic coupling parameters must meet the following constraint $\left\vert \lambda_i \right\vert < 4\pi$ due to perturbative unitarity. Taking this into account we can successfully accommodate the $125$ GeV mass for the SM like Higgs boson found at the LHC for %corresponds to 
the following VEV values:
\begin{equation}\label{eq:vevsmodel}
v_{\chi }\simeq 9.994\ \text{TeV}, \quad  v_{\eta _{2}}\simeq 244.2\ \text{GeV}, \quad v_{\rho }\simeq 29.54\ \text{GeV},
\end{equation}
which yield a mass $m_h=125.387\ \text{GeV}$ for the SM like Higgs boson. Note that the Higgs SM boson mass depends on the trilinear coupling $\varkappa$ and the quartic coupling $\lambda_8$, which is lower than its upper limit of $4\pi$ arising from perturbativity. %constrains our perturbative parameters, numerically, 
Fig. \ref{fig:kappavslambda8} displays the correlation between the trilinear scalar parameter $\varkappa$ and the quartic scalar coupling $\lambda_8$ consistent with the experimental values of the SM like Higgs boson mass. Here we have used the numerical values of the VEVs of Eq.~\eqref{eq:vevsmodel} 
%shows which scenario for the Higgs mass fit scanned the parameter space in the $\varkappa-\lambda_8$ plane, the colored bar represents the Higgs SM boson mass.  
%However, to determine the specific benchmark that reproduces the $125$ GeV mass value for the SM-like Higgs boson, the numerical values of the relevant parameters in the model are required. With the adjustment of these parameters, the numerical contributions of the physical spectrum can be determined. 
The new scalar fields introduce corrections to the phenomenological processes that arise in this model and provide more precise determinations of phenomena such as, for instance $K$ meson oscillations. In the SM-type two-photon Higgs decay constraints, where extra-charged scalar fields induce one loop level corrections to the Higgs diphoton decay. 
%At the same time, the oblique corrections are affected by the presence of extra scalar fields. 
These phenomenological processes are studied in more detail in the following sections.

\section{Meson mixings}\label{meson}

In this section we discuss the implications of our model in
the Flavour Changing Neutral Current (FCNC) interactions in the down type
quark sector. These FCNC down type quark Yukawa interactions produce $K^{0}-%
\bar{K}^{0}$, $B_{d}^{0}-\bar{B}_{d}^{0}$ and $B_{s}^{0}-\bar{B}_{s}^{0}$
meson oscillations, whose corresponding effective Hamiltonians are: 
\begin{equation}
\mathcal{H}_{eff}^{\left( K\right) }\mathcal{=}\sum_{j=1}^{3}\kappa
_{j}^{\left( K\right) }\left( \mu \right) \mathcal{O}_{j}^{\left( K\right)
}\left( \mu \right) ,
\end{equation}%
\begin{equation}
\mathcal{H}_{eff}^{\left( B_{d}\right) }\mathcal{=}\sum_{j=1}^{3}\kappa
_{j}^{\left( B_{d}\right) }\left( \mu \right) \mathcal{O}_{j}^{\left(
B_{d}\right) }\left( \mu \right) ,
\end{equation}%
\begin{equation}
\mathcal{H}_{eff}^{\left( B_{s}\right) }\mathcal{=}\sum_{j=1}^{3}\kappa
_{j}^{\left( B_{s}\right) }\left( \mu \right) \mathcal{O}_{j}^{\left(
B_{s}\right) }\left( \mu \right) ,
\end{equation}

where: 
\begin{eqnarray}
\mathcal{O}_{1}^{\left( K\right) } &=&\left( \overline{s}_{R}d_{L}\right)
\left( \overline{s}_{R}d_{L}\right) ,\hspace{0.7cm}\hspace{0.7cm}\mathcal{O}%
_{2}^{\left( K\right) }=\left( \overline{s}_{L}d_{R}\right) \left( \overline{%
s}_{L}d_{R}\right) ,\hspace{0.7cm}\hspace{0.7cm}\mathcal{O}_{3}^{\left(
K\right) }=\left( \overline{s}_{R}d_{L}\right) \left( \overline{s}%
_{L}d_{R}\right) ,  \label{op3f} \\
\mathcal{O}_{1}^{\left( B_{d}\right) } &=&\left( \overline{d}%
_{R}b_{L}\right) \left( \overline{d}_{R}b_{L}\right) ,\hspace{0.7cm}\hspace{%
0.7cm}\mathcal{O}_{2}^{\left( B_{d}\right) }=\left( \overline{d}%
_{L}b_{R}\right) \left( \overline{d}_{L}b_{R}\right) ,\hspace{0.7cm}\hspace{%
0.7cm}\mathcal{O}_{3}^{\left( B_{d}\right) }=\left( \overline{d}%
_{R}b_{L}\right) \left( \overline{d}_{L}b_{R}\right) , \\
\mathcal{O}_{1}^{\left( B_{s}\right) } &=&\left( \overline{s}%
_{R}b_{L}\right) \left( \overline{s}_{R}b_{L}\right) ,\hspace{0.7cm}\hspace{%
0.7cm}\mathcal{O}_{2}^{\left( B_{s}\right) }=\left( \overline{s}%
_{L}b_{R}\right) \left( \overline{s}_{L}b_{R}\right) ,\hspace{0.7cm}\hspace{%
0.7cm}\mathcal{O}_{3}^{\left( B_{s}\right) }=\left( \overline{s}%
_{R}b_{L}\right) \left( \overline{s}_{L}b_{R}\right) ,
\end{eqnarray}

and the Wilson coefficients take the form: 
\begin{eqnarray}
\kappa _{1}^{\left( K\right) } &=&\frac{x_{h\overline{s}_{R}d_{L}}^{2}}{%
m_{h}^{2}}+\sum_{m=1}^{5}\sum_{n=1}^{4}\left( \frac{x_{H_{m}^{0}\overline{s}_{R}d_{L}}^{2}}{%
m_{H_{m}^{0}}^{2}}-\frac{x_{A_{n}^{0}\overline{s}_{R}d_{L}}^{2}}{m_{A_{n}^{0}}^{2}}%
,\right) \\
\kappa _{2}^{\left( K\right) } &=&\frac{x_{h\overline{s}_{L}d_{R}}^{2}}{%
m_{h}^{2}}+\sum_{m=1}^{5}\sum_{n=1}^{4}\left( \frac{x_{H_{m}^{0}\overline{s}_{L}d_{R}}^{2}}{%
m_{H_{m}^{0}}^{2}}-\frac{x_{A_{n}^{0}\overline{s}_{L}d_{R}}^{2}}{m_{A_{n}^{0}}^{2}}%
\right) ,\hspace{0.7cm}\hspace{0.7cm} \\
\kappa _{3}^{\left( K\right) } &=&\frac{x_{h\overline{s}_{R}d_{L}}x_{h%
\overline{s}_{L}d_{R}}}{m_{h}^{2}}+\sum_{m=1}^{5}\sum_{n=1}^{4}\left( \frac{x_{H_{m}^{0}%
\overline{s}_{R}d_{L}}x_{H_{m}^{0}\overline{s}_{L}d_{R}}}{m_{H_{m}^{0}}^{2}}-\frac{%
x_{A_{n}^{0}\overline{s}_{R}d_{L}}x_{A_{n}^{0}\overline{s}_{L}d_{R}}}{m_{A_{n}^{0}}^{2}}%
\right) ,
\end{eqnarray}%
\begin{eqnarray}
\kappa _{1}^{\left( B_{d}\right) } &=&\frac{x_{h\overline{d}_{R}b_{L}}^{2}}{%
m_{h}^{2}}+\sum_{m=1}^{5}\sum_{n=1}^{4}\left( \frac{x_{H_{m}^{0}\overline{d}_{R}b_{L}}^{2}}{%
m_{H_{m}^{0}}^{2}}-\frac{x_{A_{n}^{0}\overline{d}_{R}b_{L}}^{2}}{m_{A_{n}^{0}}^{2}}%
\right) , \\
\kappa _{2}^{\left( B_{d}\right) } &=&\frac{x_{h\overline{d}_{L}b_{R}}^{2}}{%
m_{h}^{2}}+\sum_{m=1}^{5}\sum_{n=1}^{4}\left( \frac{x_{H_{m}^{0}\overline{d}_{L}b_{R}}^{2}}{%
m_{H_{m}^{0}}^{2}}-\frac{x_{A_{n}^{0}\overline{d}_{L}b_{R}}^{2}}{m_{A_{n}^{0}}^{2}}%
\right) , \\
\kappa _{3}^{\left( B_{d}\right) } &=&\frac{x_{h\overline{d}_{R}b_{L}}x_{h%
\overline{d}_{L}b_{R}}}{m_{h}^{2}}+\sum_{m=1}^{5}\sum_{n=1}^{4}\left( \frac{x_{H_{m}^{0}%
\overline{d}_{R}b_{L}}x_{H_{m}^{0}\overline{d}_{L}b_{R}}}{m_{H_{m}^{0}}^{2}}-\frac{%
x_{A_{n}^{0}\overline{d}_{R}b_{L}}x_{A_{n}^{0}\overline{d}_{L}b_{R}}}{m_{A_{n}^{0}}^{2}}%
\right) ,
\end{eqnarray}%
\begin{eqnarray}
\kappa _{1}^{\left( B_{s}\right) } &=&\frac{x_{h\overline{s}_{R}b_{L}}^{2}}{%
m_{h}^{2}}+\sum_{m=1}^{5}\sum_{n=1}^{4}\left( \frac{x_{H_{m}^{0}\overline{s}_{R}b_{L}}^{2}}{%
m_{H_{m}^{0}}^{2}}-\frac{x_{A_{n}^{0}\overline{s}_{R}b_{L}}^{2}}{m_{A_{n}^{0}}^{2}}%
\right) , \\
\kappa _{2}^{\left( B_{s}\right) } &=&\frac{x_{h\overline{s}_{L}b_{R}}^{2}}{%
m_{h}^{2}}+\sum_{m=1}^{5}\sum_{n=1}^{4}\left( \frac{x_{H_{m}^{0}\overline{s}_{L}b_{R}}^{2}}{%
m_{H_{m}^{0}}^{2}}-\frac{x_{A_{n}^{0}\overline{s}_{L}b_{R}}^{2}}{m_{A_{n}^{0}}^{2}}%
\right) , \\
\kappa _{3}^{\left( B_{s}\right) } &=&\frac{x_{h\overline{s}_{R}b_{L}}x_{h%
\overline{s}_{L}b_{R}}}{m_{h}^{2}}+\sum_{m=1}^{5}\sum_{n=1}^{4}\left( \frac{x_{H_{m}^{0}%
\overline{s}_{R}b_{L}}x_{H_{m}^{0}\overline{s}_{L}b_{R}}}{m_{H_{m}^{0}}^{2}}-\frac{%
x_{A_{n}^{0}\overline{s}_{R}b_{L}}x_{A_{n}^{0}\overline{s}_{L}b_{R}}}{m_{A_{n}^{0}}^{2}}%
\right) ,
\end{eqnarray}%
where the $x_{h_a\overline{d}_{cL}d_{fR}}$ effective parameters are the couplings of the physical scalar and pseudoscalar fields $h_a$ with the $d_{c,f}$-type quarks, with $h_a=h,\ H_m^0,\ A_n^0$ and $d_{c,f}=d,\; s,\; b$. Furthermore, we have used the notation of section \ref{scalar} for the physical
scalars, assuming $h$ is the lightest of the CP-even ones and
corresponds to the SM Higgs.
The $K-\bar{K}$, $B_{d}^{0}-\bar{B}_{d}^{0}$ and $B_{s}^{0}-\bar{B}_{s}^{0}$%
\ meson mass splittings read: 
\begin{equation}
\Delta m_{K}=\Delta m_{K}^{\left( SM\right) }+\Delta m_{K}^{\left( NP\right)
},\hspace{1cm}\Delta m_{B_{d}}=\Delta m_{B_{d}}^{\left( SM\right) }+\Delta
m_{B_{d}}^{\left( NP\right) },\hspace{1cm}\Delta m_{B_{s}}=\Delta
m_{B_{s}}^{\left( SM\right) }+\Delta m_{B_{s}}^{\left( NP\right) },
\label{Deltam}
\end{equation}

\begin{figure}[tbp]
\centering
\subfigure[] {\includegraphics[scale=0.4]{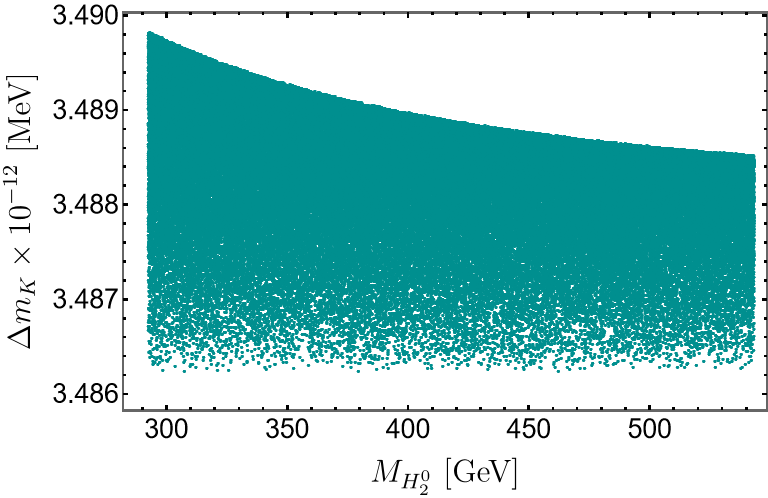}}
\quad \subfigure[]{%
\includegraphics[scale=0.41]{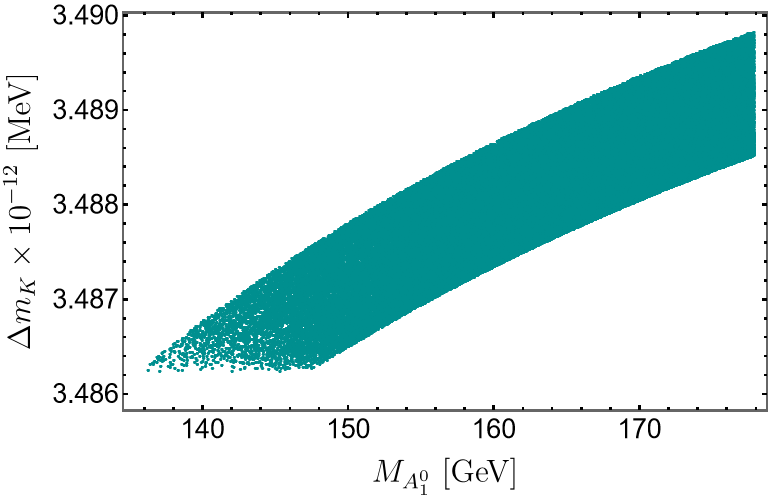}} 
\caption{Correlation a) between the $\Delta m_{K}$ mass splitting and the lightest CP even scalar mass $m_{H_2^0}$, b) between the $\Delta m_{K}$ mass splitting and the lightest CP odd scalar mass $m_{A_1^0}$.}
\label{fig:mesonmixing}
\end{figure}

where $\Delta m_{K}^{\left( SM\right) }$, $\Delta m_{B_{d}}^{\left(
SM\right) }$ and $\Delta m_{B_{s}}^{\left( SM\right) }$ correspond to the SM
contributions, while $\Delta m_{K}^{\left( NP\right) }$, $\Delta
m_{B_{d}}^{\left( NP\right) }$ and $\Delta m_{B_{s}}^{\left( NP\right) }$
are due to new physics effects. Our model predicts the following new physics
contributions for the $K-\bar{K}$, $B_{d}^{0}-\bar{B}_{d}^{0}$ and $%
B_{s}^{0}-\bar{B}_{s}^{0}$ meson mass differences: 
\begin{eqnarray}
\Delta m_{K}^{\left( NP\right) }&\simeq&\frac{8}{3}f_{K}^{2}\eta _{K}B_{K}m_{K}%
\left[ r_{2}^{\left( K\right) }\kappa _{3}^{\left( K\right) }+r_{1}^{\left(
K\right) }\left( \kappa _{1}^{\left( K\right) }+\kappa _{2}^{\left( K\right)
}\right) \right]~,\label{eq:mK} \\
\Delta m_{B_{d}}^{\left( NP\right) }&\simeq&\frac{8}{3}f_{B_{d}}^{2}\eta
_{B_{d}}B_{B_{d}}m_{B_{d}}\left[ r_{2}^{\left( B_{d}\right) }\kappa
_{3}^{\left( B_{d}\right) }+r_{1}^{\left( B_{d}\right) }\left( \kappa
_{1}^{\left( B_{d}\right) }+\kappa _{2}^{\left( B_{d}\right) }\right) \right]~, \label{eq:mBd}\\
\Delta m_{B_{s}}^{\left( NP\right) }&\simeq&\frac{8}{3}f_{B_{s}}^{2}\eta
_{B_{s}}B_{B_{s}}m_{B_{s}}\left[ r_{2}^{\left( B_{s}\right) }\kappa
_{3}^{\left( B_{s}\right) }+r_{1}^{\left( B_{s}\right) }\left( \kappa
_{1}^{\left( B_{s}\right) }+\kappa _{2}^{\left( B_{s}\right) }\right) \right]~.\label{eq:mBs}
\end{eqnarray}

Since the contribution arising from the flavor changing down type quark interaction involving the $Z^{\prime}$ gauge boson exchange is very small and subleading, the main contributions to the meson mass differences is due to the virtual exchange of additional scalar and pseudosalar fields participating in the flavor violating Yukawa interactions of the model under consideration. Using the following numerical values of the meson parameters \cite%
{Dedes:2002er,Aranda:2012bv,Khalil:2013ixa,Queiroz:2016gif,Buras:2016dxz,Ferreira:2017tvy,NguyenTuan:2020xls}%Duy:2020hhk}%
: 
\begin{eqnarray}
\left(\Delta m_{K}\right)_{\exp }&=&\left( 3.484\pm 0.006\right) \times
10^{-12}\, \mathrm{{MeV},\hspace{1.5cm}\left( \Delta m_{K}\right)_{\text{SM}}=3.483\times 10^{-12}\, {MeV}}  \notag \\
f_{K} &=&155.7\, \mathrm{{MeV},\hspace{1.5cm} B_{K}=0.85,\hspace{1.5cm}\eta_{K}=0.57,}  \notag \\
r_{1}^{\left( K\right) } &=&-9.3,\hspace{1.5cm}r_{2}^{\left(K\right) }=30.6,%
\hspace{1.5cm}m_{K}=\left(497.611\pm 0.013\right)\, \mathrm{{MeV},}
\end{eqnarray}%
\begin{eqnarray}
\left( \Delta m_{B_{d}}\right) _{\exp } &=&\left(3.334\pm 0.013\right)
\times 10^{-10}\, \mathrm{{MeV},\hspace{1.5cm}\left( \Delta m_{B_{d}}\right)
_{SM}=\left(3.653\pm 0.037\pm 0.019\right)\times 10^{-10}\, {MeV},}  \notag
\\
f_{B_{d}} &=&188\, \mathrm{{MeV},\hspace{1.5cm}B_{B_{d}}=1.26,\hspace{1.5cm}%
\eta _{B_{d}}=0.55,}  \notag \\
r_{1}^{\left( B_{d}\right) } &=&-0.52,\hspace{1.5cm}r_{2}^{\left(
B_{d}\right) }=0.88,\hspace{1.5cm}m_{B_{d}}=\left(5279.65\pm 0.12\right)\,%
\mathrm{{MeV},}
\end{eqnarray}%
\begin{eqnarray}
\left( \Delta m_{B_{s}}\right) _{\exp } &=&\left(1.1683\pm 0.0013\right)
\times 10^{-8}\, \mathrm{{MeV},\hspace{1.5cm}\left( \Delta m_{B_{s}}\right)
_{SM}=\left(1.1577\pm 0.022\pm 0.051\right) \times 10^{-8}\, {MeV},}  \notag
\\
f_{B_{s}} &=&225\, \mathrm{{MeV},\hspace{1.5cm}B_{B_{s}}=1.33,\hspace{1.5cm}%
\eta _{B_{s}}=0.55,}  \notag \\
r_{1}^{\left( B_{s}\right) } &=&-0.52,\hspace{1.5cm}r_{2}^{\left(
B_{s}\right) }=0.88,\hspace{1.5cm}m_{B_{s}}=\left(5366.9\pm 0.12\right)\, 
\mathrm{{MeV},}
\end{eqnarray}% $(B_d^0-\overline{B}_d^0)$ and $10^{-4}$, respectively

Fig. \ref{fig:mesonmixing} \textcolor{red}{(a)} and Fig. \ref{fig:mesonmixing} \textcolor{red}{(b)} show the correlations of the mass splitting $\Delta m_K$ with the mass of the lightest CP-even and CP-odd scale $m_ {H_2^0}$ and $m_ {A_1^0}$, respectively. In our numerical analysis, for the sake of simplicity, we have set the couplings of flavor-changing Yukawa neutral interactions that produce $(K^0- \overline{K}^0)$ mixings to be equal to $10^{-6}$. In addition, we have varied the masses around 20\% of their best fit-point values obtained in the analysis of the scalar sector shown in the plots of Fig. \ref{fig:scalar-sector}. As indicated in Fig. \ref{fig:mesonmixing}, our model can successfully accommodate the experimental constraints arising from $(K^0-\overline{K}^ 0) $ meson oscillations for the above specified range of parameter space. We have numerically verified that in the range of masses described above, the values obtained for the mass splittings $\Delta m_{B_d}$ and $\Delta m_{B_s}$ are consistent with the experimental data on meson oscillations for flavor violating Yukawa couplings equal to $10^{ -4}$ and $2.5\times 10^{ -4}$, respectively.

\section{Higgs di-photon decay rate}\label{di-photon}

In order to study the implications of our model in the decay of the $125\  \mathrm{GeV}$ Higgs into a photon pair, one introduces the Higgs diphoton signal strength $R_{\gamma \gamma}$, which is defined as \cite{Bhattacharyya:2014oka}:
\begin{equation}
R_{\gamma \gamma}=\frac{\sigma(p p \rightarrow h) \Gamma(h \rightarrow \gamma \gamma)}{\sigma(p p \rightarrow h)_{\text{SM}} \Gamma(h \rightarrow \gamma \gamma)_{\text{SM}}} \simeq a_{h t t}^2 \frac{\Gamma(h \rightarrow \gamma \gamma)}{\Gamma(h \rightarrow \gamma \gamma)_{\text{SM}}} .
\end{equation}

That Higgs diphoton signal strength, normalizes the $\gamma \gamma$ signal predicted by our model in relation to the one given by the SM. Here we have used the fact that in our model, single Higgs production is also dominated by gluon fusion as in the Standard Model. In the 3-3-1 model $\sigma\left( pp \rightarrow h\right)=a_{htt}^2 \sigma\left( pp \rightarrow h\right)_{\text{SM}}$, so $R_{\gamma \gamma}$ reduces to the ratio of branching ratios.\\
The decay rate for the $h\rightarrow \gamma \gamma$ process takes the form \cite{Bhattacharyya:2014oka,Logan:2014jla,Hernandez:2021uxx}:
\begin{equation}\label{diphoton}
\Gamma(h \rightarrow \gamma \gamma)=\frac{\alpha_{\text{em}}^2 m_h^3}{256 \pi^3 v^2}\left|\sum_f a_{h f f} N_C Q_f^2 F_{1 / 2}\left(\varrho_f\right)+a_{h W W} F_1\left(\varrho_W\right)+\sum_{k} \frac{C_{h H_k^{\pm} H_k^{\mp} }v}{2 m_{H_k^{\pm}}^2} F_0\left(\varrho_{H_k^{\pm}}\right)\right|^2
\end{equation}

where $\alpha_{\text{em}}$ is the fine structure constant, $N_C$ is the color factor ($N_C=3$ for quarks and $N_C=1$ for leptons) and $Q_f$ is the electric charge of the fermion in the loop. From the fermion-loop contributions we only consider the dominant top quark term. The $\varrho_i$ are the mass ratios $\varrho_i=4 M_i^2/m_h^2$ with $M_i=m_f, M_W,M_{H_k^{\pm}}$ with $k=1,2,3$. Furthermore, $C_{h H_k^{\pm} H_k^{\mp}}$ is the trilinear coupling between the SM-like Higgs and a pair of charged Higges, whereas $a_{h t t}$ and $a_{h W W}$ are the deviation factors from the SM Higgs-top quark coupling and the SM Higgs-W gauge boson coupling, respectively (in the SM these factors are unity). Such deviation factors are close to unity in our model, which is a consequence of the numerical analysis of its scalar, Yukawa and gauge sectors.

The form factors for the contributions from spin-$0$, $1/2$ and $1$ particles are:
\begin{eqnarray}
F_0(\varrho) & =& -\varrho(1-\varrho f(\varrho)),\\
F_{1 / 2}(\varrho) & =& 2\varrho(1+(1-\varrho) f(\varrho)), \\
F_1(\varrho) & =& -\left(2+3\varrho+3\varrho \left(2-\varrho\right) f(\varrho)\right),
\end{eqnarray}
with
\begin{equation}
f(\varrho)= \begin{cases}\arcsin ^2 \sqrt{\varrho^{-1}} & \text { for } \varrho \geq 1 \\ -\frac{1}{4}\left[\ln \left(\frac{1+\sqrt{1-\varrho}}{1-\sqrt{1-\varrho}}\right)-i\pi\right] ^2& \text { for } \varrho<1\end{cases}
\end{equation}

\begin{figure} 
\centering
\includegraphics[scale=0.55]{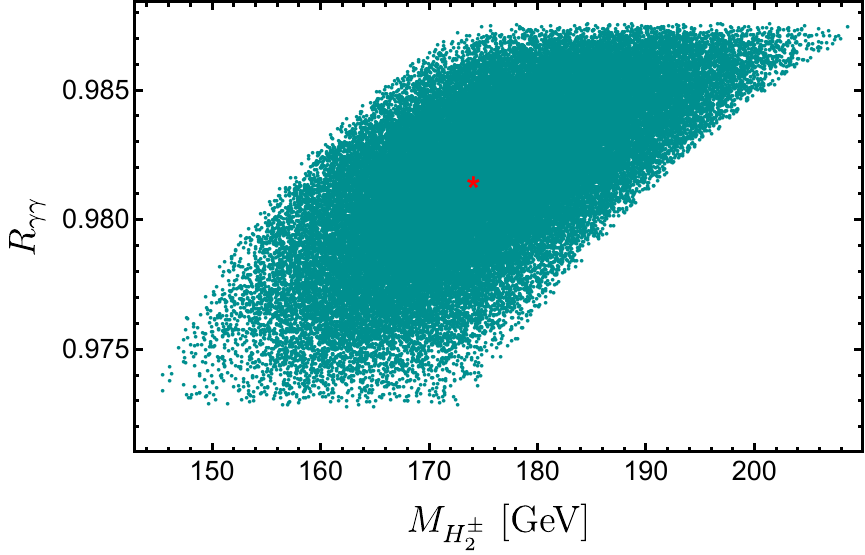}
\caption{Correlation of the Higgs di-photon signal strength with the charged scalar mass. The red star point corresponds to the best fit for $R_{\gamma \gamma}$ (see Table \ref{tab2photon}).}
\label{2photon}
\end{figure}

Table \ref{tab2photon} displays the best-fit values of the $R_{\gamma \gamma}$ ratio in comparison to the best-fit signals measured in CMS \cite{10.1007/978-981-19-2354-8_33}  and ATLAS \cite{ATLAS:2022tnm}. In this analysis, the electrically charged scalar fields play a key role in determining the value of the ratio, while the other fields have an indirect impact through the parameter space involving the VEV (vacuum expectation values) and the trilinear scalar coupling $A$, as well as some $\lambda_i$. From our numerical analysis, it follows that our model favors a Higgs diphoton decay rate lower than the SM expectation but inside the $3\sigma$ experimentally allowed range. The correlation of the Higgs diphoton signal strength with the charged scalar mass $M_{H_2^{\pm}}$ is shown in Fig. \ref{2photon}, which indicates that our model successfully accommodates the current Higgs diphoton decay rate constraints. Additionally, it should be noted that the correlation with $M_{H_1^{\pm}}$ is similar; however, the correlation is weaker than for $M_{H_3^{\pm}}$.
\begin{table}[H]
\centering
\begin{tabular}{lccc}
 \hline\hline
& Model value  & CMS & ATLAS \\ \hline
$R_{\gamma \gamma}$ & $0.982 \pm 0.08$ & $1.02_{-0.09}^{+0.11}$ & $1.04_{-0.09}^{+0.10} $  \\ 
\hline\hline
\end{tabular}
\caption{The best fit for the ratio of Higgs boson diphoton decay obtained from the model indicates a lower Higgs decay rate into two photons compared to the expectation of the Standard Model in ATLAS \cite{ATLAS:2022tnm} and CMS \cite{10.1007/978-981-19-2354-8_33} collaboration.  However, this value still falls within the $1\sigma$ experimentally allowed range.}
\label{tab2photon}
\end{table}

%%%%%%%%%%%%%%%%%%%%%%%%%%%%%%%%%%%%%%%%%%%%%%%%%%%%%%%%%%%%%%%%%%%%%%

\section{Oblique $T$, $S$ and $U$ parameters}\label{oblique}

The parameters $S$, $T$, and $U$ basically quantify the corrections to the two-point functions of gauge bosons through loop diagrams. In our case, where there are three $SU(3)_L$ scalar triplets that introduce new scalar particles, which lead to new Higgs-mediated contributions to the self-energies of gauge bosons through loop diagrams. Based on references \cite{Peskin:1991sw,Altarelli:1990zd,Barbieri:2004qk}, the parameters $S$, $T$, and $U$ can be defined as follows:
\begin{eqnarray}
 T &=& \frac{1}{\alpha_{\text{em}} M_W} \left. \left[ \Pi_{11}\left(q^2\right) - \Pi_{33} \left( q^2\right)\right] \right|_{q=0} \\
 S & = &-\left.\frac{4  c_W s_W}{ \alpha_{\text{em}}} \frac{d}{d q^2} \Pi_{30}\left(q^2\right)\right|_{q^2=0},\\
 U &= & \frac{4 s_W }{\alpha_{\text{em}} } \left. \frac{d}{d q^2}\left[\Pi_{11}\left(q^2\right)-\Pi_{33}\left(q^2\right)\right]\right|_{q^2=0} ,
 \end{eqnarray}
with $s_W=\sin \theta_W$ and $c_W=\cos\theta_W$, where $\theta_W$ is the electroweak mixing angle, the quantity $\Pi_{ij} \left(q \right)$ is defined in terms of the vacuum-polarization tensors
\begin{equation}
\Pi_{ij} ^{\mu \nu} \left(q^2\right) = g^{\mu \nu} \Pi_{ij}\left(q^2\right)-i  q^{\mu} q^{\nu}  \Delta_{ij}\left(q^2\right),
\end{equation}
where $i,j=0,1,3$ for the $B$, $W_1$ and $W_3$ bosons respectively, or possibly $i,j=W,Z,\gamma$. If the new physics enters at the TeV scale, the effect of the theory will be well-described by an expansion up to linear order in $q^2$ for $\Pi_{i j} \left(q^2\right)$ as presented in reference \cite{Peskin:1991sw}. 

For our 331 model, the scalar fields arising from the $SU(3)_L$ triplets $\eta_k$ ($k=1,2$) and $\rho$ provide the dominant contributions to the %directly contribute to the 
new physics values of the 
$T$, $S$, and $U$ oblique parameters, as they couple with the $W$ and $Z$, and then we must take into account the scalar mixing angles. 
%matrices presented in Eq. (\ref{mixingangles}). 
We can calculate the parameters considering %low energy assumption to simplify our analysis, where 
that the low energy effective field theory below the scale of spontaneos breaking of the $SU(3)_L\times U(1)_X\times U(1)_{L_g}$ symmetry corresponds to a three Higgs doublet model (3HDM), where the three Higgs doublets arise from the $\eta_1$, $\eta_2$ and $\rho$ $SU(3)_L$ scalar triplets. Then, following these considerations, in the above described low energy limit scenario, the leading contributions to the oblique $T$, $S$ and $U$ parameters take the form \cite{Long:1999bny,Grimus:2007if,Grimus:2008nb,CarcamoHernandez:2015smi,Long:2018dun,CarcamoHernandez:2022vjk,Batra:2022arl}:
%if we restrict ourselves to the limit scenario where three flavor triplets can be considered a three Higgs doublet model (3HDM). 
%For a N-HDM theory \cite{Long:1999bny,Long:2018dun,CarcamoHernandez:2022vjk,Batra:2022arl}, we can generalize the following expressions considering our model:
\begin{eqnarray}
T & \simeq & t_0 \left[ 
\sum_{a=1}^{2}\sum_{k=1}^2\left[\left(R_C\right)_{a k}\right]^2 m_{H_k^{ \pm}}^2
+\sum_{a=1}^{2} \sum_{i=1}^2 \sum_{j=1}^2\left[ \left(R_H\right)_{a i} \left(R_A\right)_{a j}\right]^2 F\left(m_{H_i^0}^2, m_{A_j^0}^2\right)
\right. ,\\
\nonumber & & \left.- \sum_{a=1}^{2}\sum_{i=1}^2 \sum_{k=1}^2 \left\lbrace  \left[\left(R_H\right)_{a i}\left(R_C\right)_{a k}\right]^2 F\left(m_{H_i^0}^2, m_{H_k^{ \pm}}^2\right)+
\left[\left(R_A\right)_{a i}\left(R_C\right)_{a k}\right]^2 F\left(m_{A_i^0}^2, m_{H_k^{ \pm}}^2\right)\right\rbrace \right]  \\
\nonumber &&\\
S & \simeq & \frac{1}{12 \pi} \sum_{i=1}^2 \sum_{j=1}^2 \sum_{k=1}^2 \left[\left(R_H\right)_{k i}\left(R_A\right)_{k j}\right]^2 K\left(m_{H_i^0}^2, m_{A_j^0}^2, m_{H_k^{ \pm}}^2\right) ,\\
\nonumber &&\\
U &\simeq & -S+\sum_{a=1}^{2}\sum_{i=1}^2 \sum_{k=1}^2  \left\lbrace \left[ \left(R_A\right)_{a i}\left(R_C\right)_{a k}\right]^2 G\left(m_{A_i^0}^2, m_{H_k^{ \pm}}^2\right)  +\left[\left(R_H\right)_{a i}\left(R_C\right)_{a k}\right]^2 G\left(m_{H_i^0}^2, m_{H_k^{ \pm}}^2\right) \right\rbrace ,
\end{eqnarray}
where $t_0=\left( 16\pi^2 v^2 \alpha_{\text{em}}\left(M_Z\right) \right)^{-1}$ and $R_C$, $R_H$, $R_A$ are the mixing matrices for the charged scalar fields, neutral scalar and pseudoscalars, respectively presented in the Sec. \ref{scalarsector}. Furthermore, the following loop functions $F\left(m_1^2, m_2^2\right)$, $G\left(m_1^2, m_2^2\right)$ and $K\left(m_1^2, m_2^2, m_3^2\right)$ were introduced in \cite{CarcamoHernandez:2022vjk}:
\begin{eqnarray}
F\left(m_1^2, m_2^2\right) &=& \frac{m_1^2 m_2^2}{m_1^2-m_2^2} \ln \left(\frac{m_1^2}{m_2^2}\right),\\
G\left(m_1^2, m_2^2\right)&=&\frac{-5 m_1^6+27 m_1^4 m_2^2-27 m_1^2 m_2^4+6\left(m_1^6-3 m_1^4 m_2^2\right) \ln \left(\frac{m_1^2}{m_2^2}\right)+5 m_2^6}{6\left(m_1^2-m_2^2\right)^3},\\
K\left(m_1^2, m_2^2, m_3^2\right) &= & \frac{1}{\left(m_2^2-m_1^2\right)^3}\left\{m_1^4\left(3 m_2^2-m_1^2\right) \ln \left(\frac{m_1^2}{m_3^2}\right)-m_2^4\left(3 m_1^2-m_2^2\right) \ln \left(\frac{m_2^2}{m_3^2}\right)\right. \\
\nonumber & & \left.-\frac{1}{6}\left[27 m_1^2 m_2^2\left(m_1^2-m_2^2\right)+5\left(m_2^6-m_1^6\right)\right]\right\}.
\end{eqnarray}

Besides that, the experimental limits for $S$, $T$, and $U$ are given in ref \cite{Workman:2022ynf}:
\begin{eqnarray}
T_{\text{exp}} &= & 0.03 \pm 0.12\\
S_{\text{exp}}  &=& -0.02 \pm 0.1\\
U_{\text{exp}}  &= & 0.01 \pm 0.11
\end{eqnarray}

\begin{figure} 
\centering
\subfigure[]{\includegraphics[scale=.5]{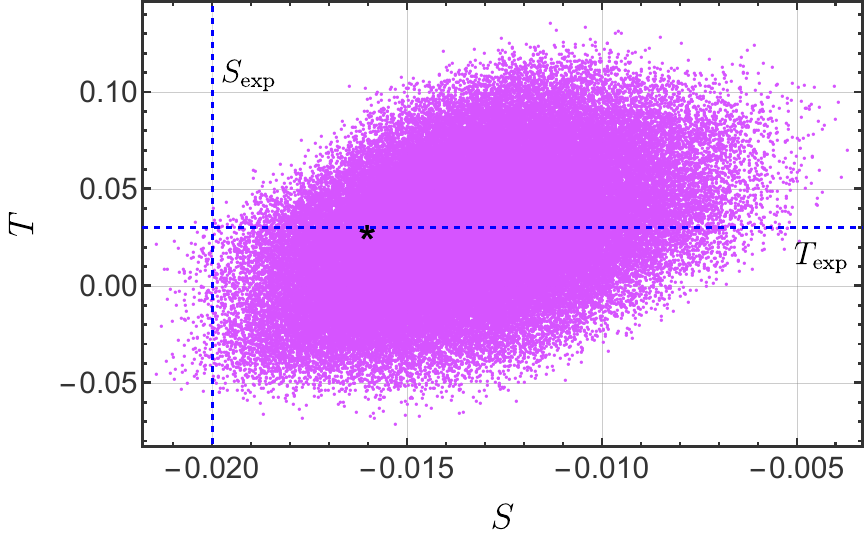}}
\subfigure[]{\includegraphics[scale=.5]{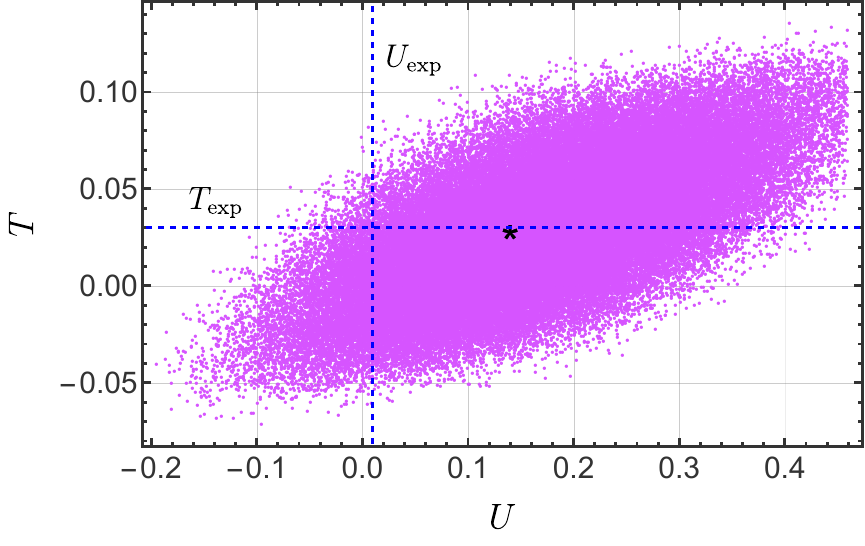}}   
\subfigure[]{\includegraphics[scale=.5]{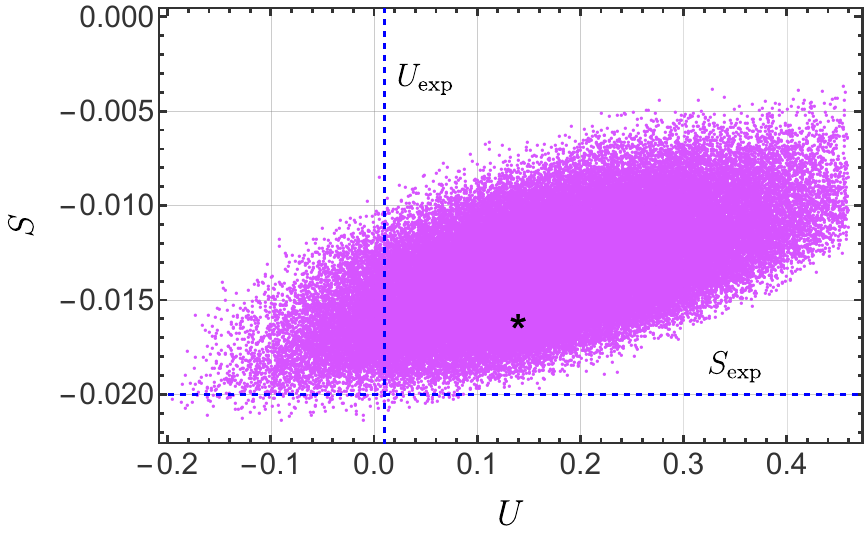}}
\caption{Correlation between the oblique parameters, the blue dashed line represents the central value of the Ref.~\cite{Workman:2022ynf} and the star corresponds to the best fit of this model.}
\label{oblicue}
\end{figure}

From the numerical analysis, the $S_4$ flavored $331$ model has restricted parameters because the determination of the new physics from $S$, $T$, and $U$ is determined by the physical masses of the model, also within the limit indicated by the oblique parameters present some correlation shown in Fig.~\ref{oblicue}, we see that the evolution of parameter space is adjusted within the $1 \sigma$ experimental range.
%which corresponds to the overlying region.
The figures \ref{oblicue} \textcolor{red}{(a)} and \textcolor{red}{(c)} of dispersion involving the $U$ parameter is produced with values larger than the central one, despite this, the values of $U$ fit within the range of $ 1 \sigma$ and the statistical discrepancy is minimal. In the case of the $S$ value, due to the large uncertainty value, the value fits more naturally, as shown in the figure \ref{oblicue} \textcolor{red}{(b)}.

%Our analysis shows consistency with the current experimental limits for our three flavor triplet models, considering the limit at 3HDM, consistent with the lighter scalar masses of the models.
Our analysis shows that our model allows a successfull fit for the oblique $S$, $T$ and $U$ parameters, consistent with their current experimental limits.
%However, the experimental constraints on the $T$, $S$, and $U$ parameters of the model parameter space impose some non-trivial restrictions. 
The obtained best fit point values for the oblique $S$, $T$ and $U$ parameters in our model are: 
\begin{eqnarray}
T &= & 0.029 \pm 0.009,\\
S &= & -0.016\pm 0.006, \\
U &= & 0.14\pm 0.04.
\end{eqnarray}

Our results also suggest that the model favors a larger value for $U$ within statistical uncertainty. While the values for $S$ and $T$ fit within the relative error $0.2$ and $0.1$, respectively.

\section{Conclusions}\label{conclu}
We have constructed a 3-3-1 model where the $SU(3)_C\times SU(3)_L\times U(1)_X$ symmetry is enlarged by the inclusion of the spontaneously broken $U(1)_{L_g}\times S_4\times Z_4\times Z_4^{\prime}\times Z_2$ symmetry group. In our theory under consideration, the observed SM charged fermion masses and the quark mixing pattern arise from the spontaneous breaking of discrete symmetries and the tiny active neutrino masses are generated from an inverse seesaw  mechanism. Our proposed model leads to a successful fit to quark and lepton masses, mixing angles, and CP phases, whose obtained values are consistent with the experimental data within the $3\sigma$ range. The symmetries of the model give rise to correlations between the mixing angles and the Jarlskog invariant for the quark sector. Regarding the lepton sector, our model predicts a diagonal SM charged lepton mass matrix, thus implying that the leptonic mixing only arises from the neutrino sector, where correlations between the leptonic mixing angles and the leptonic CP violating phase are obtained.     %It is possible to reproduce and adjust the fermionic sector, obtaining values of the masses and mixtures within the range of $3\sigma$ of the experimental values. 
%In addition, we obtain correlations between the mixing angles and the Jarlskog invariant for the quark sector, while in the charged lepton sector, we obtain a diagonal mass matrix due to the symmetry $Z_4^{\prime}$, so the mixture of the lepton sector comes exclusively from the neutrino sector, in which we have generated the small masses of the charged leptons. 
%Light-active neutrinos through an inverse Seesaw mechanism, where the values of the masses and mixtures are in the experimental range of $3\sigma$, and we obtain correlations between the mixing angles and the CP violation phase. 
In addition, flavor-changing neutral current interactions in the quark sector mediated by CP even and odd CP scalars give rise to meson oscillations, such as the $(K^0-\overline{K}^0)$ mixing, whose experimental constraints are successfully satisfied for an appropriate region of the parameter space. The theory under consideration is consistent with the masses and mixings of SM fermions as well as with the constraints arising from $(K^0-\overline{K}^0)$ and $(B_{d,s}^0-\overline {B}_{d,s}^0)$ meson oscillations. The charged scalars of our model provide the new physics contribution to the Higgs diphoton decay rate. In contrast, the rest of the scalar fields have an indirect influence involving the VEVs and \textcolor{black}{the trilinear scalar coupling $\varkappa$}, where our model favors a $R_{\gamma\gamma}$ value lower than SM, however, is within the $3\sigma$ experimentally allowed range measured by the CMS and ATLAS collaborations. The extra scalar fields of our model produce radiative corrections to the oblique parameters $S$, $T$, and $U$, where the numerical analysis yield correlations between these parameters and, in addition, their obtained values are within the $1 \sigma$ experimentally allowed range.

\section*{Acknowledgments}
A.E.C.H is supported by ANID-Chile FONDECYT 1210378, ANID PIA/APOYO AFB230003 and Milenio-ANID-ICN2019\_044 and ANID Programa de Becas Doctorado Nacional code 21212041.
\appendix

\section{The product rules of the $S_{4}$ discrete group}

\label{S4}The $S_{4}$ is the smallest non abelian group having doublet,
triplet and singlet irreducible representations. $S_{4}$ is the group of
permutations of four objects, which includes five irreducible
representations, i.e., $\mathbf{1,1^{\prime },2,3,3^{\prime }}$ fulfilling
the following tensor product rules \cite{Ishimori:2010au}:
$$
\begin{gathered}
\mathbf{3} \otimes \mathbf{3}=\mathbf{1} \oplus \mathbf{2} \oplus \mathbf{3} \oplus \mathbf{3}^{\prime}, \quad \mathbf{3}^{\prime} \otimes \mathbf{3}^{\prime}=\mathbf{1} \oplus \mathbf{2} \oplus \mathbf{3} \oplus \mathbf{3}^{\prime}, \quad \mathbf{3} \otimes \mathbf{3}^{\prime}=\mathbf{1}^{\prime} \oplus \mathbf{2} \oplus \mathbf{3} \oplus \mathbf{3}^{\prime}, \\
\mathbf{2} \otimes \mathbf{2}=\mathbf{1} \oplus \mathbf{1}^{\prime} \oplus \mathbf{2}, \quad \mathbf{2} \otimes \mathbf{3}=\mathbf{3} \oplus \mathbf{3}^{\prime}, \quad \mathbf{2} \otimes \mathbf{3}^{\prime}=\mathbf{3}^{\prime} \oplus \mathbf{3} \\
\mathbf{3} \otimes \mathbf{1}^{\prime}=\mathbf{3}^{\prime}, \quad \mathbf{3}^{\prime} \otimes \mathbf{1}^{\prime}=\mathbf{3}, \quad \mathbf{2} \otimes \mathbf{1}^{\prime}=\mathbf{2}
\end{gathered}
$$
Explicitly, the basis used in this paper corresponds to Ref. \cite{Ishimori:2010au} and results in

\begin{eqnarray}
\begin{pmatrix}
a_{1} \\ 
a_{2} \\ 
a_{3}%
\end{pmatrix}_{\mathbf{3}}\otimes 
\begin{pmatrix}
b_{1} \\ 
b_{2} \\ 
b_{3}%
\end{pmatrix}_{\mathbf{3}}& = & (a_{1}b_{1}+a_{2}b_{2}+a_{3}b_{3})_{\mathbf{1}_{1}}\oplus 
\begin{pmatrix}
\frac{1}{\sqrt{2}}(a_{2}b_{2}-a_{3}b_{3}) \\ 
\frac{1}{\sqrt{6}}(-2a_{1}b_{1}+a_{2}b_{2}+a_{3}b_{3})%
\end{pmatrix}_{\mathbf{2}}  \notag \\
&  & \oplus\begin{pmatrix}
a_{2}b_{3}+a_{3}b_{2} \\ 
a_{1}b_{3}+a_{3}b_{1} \\ 
a_{1}b_{2}+a_{2}b_{1}%
\end{pmatrix}%
_{\mathbf{3}}\oplus 
\begin{pmatrix}
a_{3}b_{2}-a_{2}b_{3} \\ 
a_{1}b_{3}-a_{3}b_{1} \\ 
a_{2}b_{1}-a_{1}b_{2}%
\end{pmatrix}_{\mathbf{3'}\ ,}
\end{eqnarray}

\begin{eqnarray}
\begin{pmatrix}
a_{1} \\ 
a_{2} \\ 
a_{3}%
\end{pmatrix}_{\mathbf{3'}}\otimes 
\begin{pmatrix}
b_{1} \\ 
b_{2} \\ 
b_{3}%
\end{pmatrix}_{\mathbf{3'}}& = & (a_{1}b_{1}+a_{2}b_{2}+a_{3}b_{3})_{\mathbf{1}_{1}}\oplus 
\begin{pmatrix}
\frac{1}{\sqrt{2}}(a_{2}b_{2}-a_{3}b_{3}) \\ 
\frac{1}{\sqrt{6}}(-2a_{1}b_{1}+a_{2}b_{2}+a_{3}b_{3})%
\end{pmatrix}_{\mathbf{2}}  \notag \\
& & \oplus 
\begin{pmatrix}
a_{2}b_{3}+a_{3}b_{2} \\ 
a_{1}b_{3}+a_{3}b_{1} \\ 
a_{1}b_{2}+a_{2}b_{1}%
\end{pmatrix}_{\mathbf{3}}\oplus 
\begin{pmatrix}
a_{3}b_{2}-a_{2}b_{3} \\ 
a_{1}b_{3}-a_{3}b_{1} \\ 
a_{2}b_{1}-a_{1}b_{2}%
\end{pmatrix}_{\mathbf{3'}\ ,}
\end{eqnarray}

\begin{eqnarray}
\begin{pmatrix}
a_{1} \\ 
a_{2} \\ 
a_{3}%
\end{pmatrix}%
_{\mathbf{3}}\otimes 
\begin{pmatrix}
b_{1} \\ 
b_{2} \\ 
b_{3}%
\end{pmatrix}%
_{\mathbf{3'}}& =& (a_{1}b_{1}+a_{2}b_{2}+a_{3}b_{3})_{\mathbf{1}%
_{2}}\oplus 
\begin{pmatrix}
\frac{1}{\sqrt{6}}(2a_{1}b_{1}-a_{2}b_{2}-a_{3}b_{3}) \\ 
\frac{1}{\sqrt{2}}(a_{2}b_{2}-a_{3}b_{3})%
\end{pmatrix}_{\mathbf{2}}  \notag \\
&  & \oplus 
\begin{pmatrix}
a_{3}b_{2}-a_{2}b_{3} \\ 
a_{1}b_{3}-a_{3}b_{1} \\ 
a_{2}b_{1}-a_{1}b_{2}%
\end{pmatrix}%
_{\mathbf{3}}\oplus 
\begin{pmatrix}
a_{2}b_{3}+a_{3}b_{2} \\ 
a_{1}b_{3}+a_{3}b_{1} \\ 
a_{1}b_{2}+a_{2}b_{1}%
\end{pmatrix}
_{\mathbf{3'}\ .}
\end{eqnarray}

\begin{eqnarray}
\begin{pmatrix}
a_{1} \\ 
a_{2}%
\end{pmatrix}_{\mathbf{2}}\otimes 
\begin{pmatrix}
b_{1} \\ 
b_{2}%
\end{pmatrix}_{\mathbf{2}}& = & (a_{1}b_{1}+a_{2}b_{2})_{\mathbf{1}}\oplus
(-a_{1}b_{2}+a_{2}b_{1})_{\mathbf{1'}}\oplus 
\begin{pmatrix}
a_{1}b_{2}+a_{2}b_{1} \\ 
a_{1}b_{1}-a_{2}b_{2}%
\end{pmatrix}%
_{\mathbf{2}\ ,} \\
\begin{pmatrix}
a_{1} \\ 
a_{2}%
\end{pmatrix}_{\mathbf{2}}\otimes 
\begin{pmatrix}
b_{1} \\ 
b_{2} \\ 
b_{3}%
\end{pmatrix}%
_{\mathbf{3}}& = &
\begin{pmatrix}
a_{2}b_{1} \\ 
-\frac{1}{2}(\sqrt{3}a_{1}b_{2}+a_{2}b_{2}) \\ 
\frac{1}{2}(\sqrt{3}a_{1}b_{3}-a_{2}b_{3})%
\end{pmatrix}%
_{\mathbf{3}}\oplus 
\begin{pmatrix}
a_{1}b_{1} \\ 
\frac{1}{2}(\sqrt{3}a_{2}b_{2}-a_{1}b_{2}) \\ 
-\frac{1}{2}(\sqrt{3}a_{2}b_{3}+a_{1}b_{3})%
\end{pmatrix}%
_{\mathbf{3'}\ ,} \\
\begin{pmatrix}
a_{1} \\ 
a_{2}%
\end{pmatrix}%
_{\mathbf{2}}\otimes 
\begin{pmatrix}
b_{1} \\ 
b_{2} \\ 
b_{3}%
\end{pmatrix}%
_{\mathbf{3'}}& =&
\begin{pmatrix}
a_{1}b_{1} \\ 
\frac{1}{2}(\sqrt{3}a_{2}b_{2}-a_{1}b_{2}) \\ 
-\frac{1}{2}(\sqrt{3}a_{2}b_{3}+a_{1}b_{3})%
\end{pmatrix}%
_{\mathbf{3}}\oplus 
\begin{pmatrix}
a_{2}b_{1} \\ 
-\frac{1}{2}(\sqrt{3}a_{1}b_{2}+a_{2}b_{2}) \\ 
\frac{1}{2}(\sqrt{3}a_{1}b_{3}-a_{2}b_{3})%
\end{pmatrix}_{\mathbf{3'}\ ,} 
\end{eqnarray}

\section{The scalar potential for a $S_4$ doublet and triplet}
\label{ScalarS4doubletandtriplet}
The scalar potential for a $S_4$ doublet $\eta$ is given by
\begin{equation} \label{VD}
V_D=-g_{\eta}^2 \left( \eta \eta^*\right)_\textbf{1}+k_1 \left( \eta \eta^*\right)_\textbf{1}\left( \eta \eta^*\right)_\textbf{1} +k_2 \left( \eta \eta^*\right)_{\textbf{1}^{\prime}}\left( \eta \eta^*\right)_{\textbf{1}^{\prime}} + k_3   \left( \eta \eta^*\right)_\textbf{2} \left( \eta \eta^*\right)_\textbf{2}.
\end{equation} 

The components of this expression are characterized by the presence of four unconstrained parameters: a bilinear coupling and three quartic couplings. From the scalar potential minimization: 
\begin{equation}
    \begin{aligned}
        \frac{\partial  \langle V_D \rangle}{\partial v_{\eta_2}} = & \  -2 g_{\eta}^2 v_{\eta_2} +4 k_1 v_{\eta_2}^3+4 k_3 v_{\eta_2}^3 = 0,
    \end{aligned}
\end{equation}
we obtain the parameter $g_{\eta}$ as a function of the other three parameters, i.e.
\begin{equation}
    \begin{aligned}
        g_{\eta}^2 = & \ 2 \left(k_1+k_3 \right) v_{\eta_2}^2,
    \end{aligned}
\end{equation}
with $k_1, \ k_3 \in \mathbb{R}$.
On the other hand, the condition of having a global minimum:
\begin{equation}
    \begin{aligned}
        \frac{\partial^2 \langle V_D \rangle}{\partial v_{\eta_2}^2}  = & \  -g_{\eta}^2 + 6 v_{\eta_{2}}^{2} (k_1 + k_3) > 0,
    \end{aligned}
\end{equation}
gives rise to the followingfollowing inequality:
\begin{equation}
    \begin{aligned}
        g_{\eta}^2 < & \ 6 v_{\eta_{2}}^{2} (k_1 + k_3),
    \end{aligned}
\end{equation}.

This result suggests that the VEV pattern of the $S_4$ doublet $\eta$ as presented in Eq. (\ref{veveta}), is consistent with a global minimum of the scalar potential of Eq. (\ref{VD}), in a large parameter space region.

%\blue{By means of the scalar potential minimization second condition we have:
%\begin{equation*}
%\frac{\partial}{\partial v_{\eta_2}} \left( \frac{\partial \langle V_D \rangle}{\partial v_{\eta_2}} \right) = \frac{\partial}{\partial v_{\eta_2}} \left( -2 g_{\eta}^2 v_{\eta_2} + 4 k_1 v_{\eta_2}^3 + 4 k_3 v_{\eta_2}^3 \right) > 0,
%\end{equation*}
%\begin{equation*}
%-2 g_{\eta}^2 + 4 k_1 (3 v_{\eta_2}^2) + 4 k_3 (3 v_{\eta_2}^2) > 0 \ \  \rightarrow \ \  2 \{ -g_{\eta}^2 + 6 v_{\eta_{2}}^{2} (k_1 + k_3) \} > 0,
%\end{equation*}
%\begin{equation*}
%-g_{\eta}^2 + 6 v_{\eta_{2}}^{2} (k_1 + k_3) > 0,
%\end{equation*}
%}

%While the scalar triplet potential $S_1$ under the discrete group $S_4$ is written as:
%\begin{eqnarray} \label{VT}
%V_T &= & -g_{S_1}^2 \left( S_1 S_1^*\right)_\textbf{1}+k_1 \left( S_1 S_1^*\right)_\textbf{1}\left( S_1 S_1^*\right)_\textbf{1} +k_2 \left( S_1 S_1^*\right)_{\textbf{3}}\left( S_1 S_1^*\right)_{\textbf{3}}+ k_3 \left( S_1 S_1^*\right)_{\textbf{3}^{\prime}}\left( S_1 S_1^*\right)_{\textbf{3}^{\prime}} \\
%\nonumber & &+ k_4 \left( S_1 S_1^*\right)_\textbf{2} \left( S_1 S_1^*\right)_\textbf{2} +H.c.
%\end{eqnarray}  
The relevant terms determining the VEV directions of any $S_4$ scalar triplet are:
\begin{eqnarray}
V_T &= & -g_{\Omega}^2 \left( \Omega \Omega^*\right)_\textbf{1}+k_1 \left( \Omega \Omega^*\right)_\textbf{1}\left( \Omega \Omega^*\right)_\textbf{1} +k_2 \left( \Omega \Omega^*\right)_{\textbf{3}}\left( \Omega \Omega^*\right)_{\textbf{3}}+ k_3 \left( \Omega \Omega^*\right)_{\textbf{3}^{\prime}}\left( \Omega \Omega^*\right)_{\textbf{3}^{\prime}} \\
\nonumber & &+ k_4 \left( \Omega \Omega^*\right)_\textbf{2} \left( \Omega \Omega^*\right)_\textbf{2}
+H.c.
\label{VT}
\end{eqnarray} 

where $\Omega=S_e,S_{\mu},S_{\tau},\Phi,\xi$.

The part of the scalar potential for each $S_4$ scalar triplet has five free parameters: one bilinear and
four quartic couplings. The minimization conditions of the scalar potential for a $S_4$ triplet yield the
following relations:

\begin{eqnarray} \label{vevS4}
\nonumber \frac{\partial \langle V_T \rangle}{\partial v_{\Omega_1}}   & = & -2g_{\Omega}^2 v_{\Omega_1} +8 k_2 v_{\Omega_1}  \left(v_{\Omega_2}^2 +v_{\Omega_3}^2 \right) + 4k_1 v_{\Omega_1} \left( v_{\Omega_1}^2 + v_{\Omega_2}^2 + v_{\Omega_3}^2 \right) - \frac{4}{3} k_4 v_{\Omega_1} \left( -2v_{\Omega_1}^2 + v_{\Omega_2}^2 + v_{\Omega_3}^2 \right)  \\
 \nonumber & = & 0, \\
 \nonumber & & \\
\frac{\partial \langle V_T \rangle}{\partial v_{\Omega_2}}   & = & -2g_{\Omega}^2 v_{\Omega_2} + 8 k_2 v_{\Omega_2} \left( v_{\Omega_1}^2 + v_{\Omega_3}^2  \right)+ 4 k_1 v_{\Omega_2} \left( v_{\Omega_1}^2 + v_{\Omega_2}^2 + v_{\Omega_3}^2 \right) +2k_4  v_{\Omega_2}  \left( v_{\Omega_2}^2 - v_{\Omega_3}^2\right) \\
\nonumber & & + \frac{2}{3} k_4 v_{\Omega_2} \left(-2v_{\Omega_1}^2 +v_{\Omega_2}^2 + v_{\Omega_3}^2 \right) \\
 \nonumber & = & 0, \\
\nonumber  & & \\
\nonumber \frac{\partial \langle V_T \rangle}{\partial v_{\Omega_3}}   & = & -2g_{\Omega}^2 v_{\Omega_3}
+8 k_2 v_{\Omega_3}   \left( v_{\Omega_1}^2 + v_{\Omega_2}^2 \right) + 4k_1 v_{\Omega_3} \left( v_{\Omega_1}^2 + v_{\Omega_2}^2 + v_{\Omega_3}^2 \right)- 2k_4  v_{\Omega_3}  \left( v_{\Omega_2}^2 - v_{\Omega_3}^2\right) \\
\nonumber & & + \frac{2}{3} k_4v_{\Omega_3} \left(-2v_{\Omega_1}^2 +v_{\Omega_2}^2 + v_{\Omega_3}^2 \right) \\
 \nonumber & = & 0.
\end{eqnarray}
whereas the condition of having a global minimum gives rise to the following inequalities:
\begin{equation}
\begin{aligned}
\frac{\partial^2 \langle V_T \rangle}{\partial v_{\Omega_1}^2} = & \  -2 g_{\Omega}^2 + 8 k_{2} \left( v_{\Omega_{2}}^{2} + v_{\Omega_{3}}^{2} \right) + 4 k_{1} \left( 3 v_{\Omega_{1}}^{2} + v_{\Omega_{2}}^{2} + v_{\Omega_{3}}^{2} \right) + \frac{4}{3} k_{4} \left( 6 v_{\Omega_{1}}^{2} - v_{\Omega_{2}}^{2}- v_{\Omega_{3}}^{2} \right) > 0, \\
 & \\
\frac{\partial^2 \langle V_T \rangle}{\partial v_{\Omega_2}^2} = & -2 g_{\Omega}^2 + 8 k_{2} \left( v_{\Omega_{1}}^{2} + v_{\Omega_{3}}^{2} \right) + 4 k_{1} \left( v_{\Omega_{1}}^{2} + 3 v_{\Omega_{2}}^{2} + v_{\Omega_{3}}^{2} \right) + 2 k_{4} \left( 3 v_{\Omega_{2}}^{2} - v_{\Omega_{3}}^{2} \right) + \\
& - \frac{4}{3} k_{4} \left(  v_{\Omega_{1}}^{2} -6 v_{\Omega_{2}}^{2} + v_{\Omega_{3}}^{2} \right) > 0, \\
 & \\
\frac{\partial^2 \langle V_T \rangle}{\partial v_{\Omega_3}^2} = & -2 g_{\Omega}^2 + 8 k_{2} \left( v_{\Omega_{1}}^{2} + v_{\Omega_{2}}^{2} \right) + 4 k_{1} \left( v_{\Omega_{1}}^{2} + v_{\Omega_{2}}^{2} + 3 v_{\Omega_{3}}^{2} \right) - 2 k_{4} \left( v_{\Omega_{2}}^{2} - 3 v_{\Omega_{3}}^{2} \right) \\ 
& - \frac{4}{3} k_{4} \left(  v_{\Omega_{1}}^{2} + v_{\Omega_{2}}^{2}-6 v_{\Omega_{3}}^{2} \right) > 0,
\end{aligned}
\end{equation}

From the scalar potential minimization equations. From the first minimization condition we get:
%, we obtain the following \blue{relation for the parameter $g_{S_{e}}$}:
%For $S_e$, we obtain from the scalar potential minimization equations it follows that:
\begin{eqnarray}
g_{S_k}^2 & = & \frac{\left(3k_1 + 2k_4 \right)}{3}v_{S_k}^2, \\
g_{\Phi}^2 & = &\frac{2  \left(2 e^{4 i \theta } \left(6 \left(k_1+k_2\right)+k_4\right) r_1^4+4 e^{2 i\theta } \left(3 k_1+6 k_2-k_4\right) r_1^2+3 k_1+2 k_4\right)}{3+6 e^{2 i \theta } r_1^2} v_\Phi^2,\\
g_{\xi}^2&=& \frac{2  \left(2 k_4 \left(r_2^2-1\right){}^2+3 k_1 \left(r_2^2+2\right){}^2+12 k_2 \left(2
   r_2^2+1\right)\right)}{3 \left(r_2^2+2\right)} v_\xi^2,
\end{eqnarray}
where $k=e,\mu,\tau$.
This shows that the VEV configuration of the $S_4$ triplet $S_e$
given in the equations \eqref{eq:vev-lep} and \eqref{eq:vev-lep2}, is in accordance with the scalar potential minimization condition of Eq. \eqref{vevS4}. The other $S_4$ triplets in our model are also consistent with the scalar potential minimization conditions, which can be demonstrated using the same procedure described in this appendix.
These results show that the VEV directions for the $S_4$ triplets $S_e$, $S_{\mu}$, $S_{\tau}$, $\Phi$, $\xi$ are consistent with a global minimum of the scalar potential for a large region of parameter space.

\section{\label{apex:matrixscalar}Mass squared mastrices for scalar sector}

In the de Born level analysis of the mass spectrum of the Higgs bosons and the physical basis of scalar particles, we must construct the scalar mass matrices of the model. Then substituting the equations of constraintst Eq.~(\ref{constraints}), and Eq.~\eqref{triplet} into the scalar potential Eq.~\eqref{Vhiggs}, the square mass matrices are determined, calculating the second derivatives of the potential
\begin{equation}
\left(M^2_{\phi^{\pm}}\right)_{ij}=\left.\frac{\partial V}{\partial \phi^+_i \partial \phi_j^-}\right\vert_{\left\langle \phi^{\pm}\right\rangle }, \quad \quad
\left(M^2_{\phi}\right)_{ij}=\left.\frac{\partial V}{\partial \phi_i \partial \phi_j}\right\vert_{\left\langle \phi\right\rangle },
\end{equation}

where $\phi^{\pm}=\chi_2^{+},\rho_1^{+},\rho_3^{+},\eta_{21}^{+},\eta_{22}^{+} ,\chi_2^{-},\rho_1^{-},\rho_3^{-},\eta_{21}^{-},\eta_{22}^{-}$, for charged fields and $\phi =$ $\xi_\chi,\xi_\rho,\xi_{\eta_1},\xi_{\eta_2},\zeta_{\chi},$ $\zeta_{\rho},$ $\zeta_{\eta_1 },$ $\zeta_{\eta_2},$ $\chi_1^0,\eta_{31}^0,\eta_{32}^0$, for neutral scalar fields. Due to the symmetry of our models, the matrices of the CP odd and CP even sectors contain two diagonal blocks, a situation similar to that presented in other works on 3-3-1 models \cite{CarcamoHernandez:2019iwh}, however, our results differ in higher dimension matrices due to the extra intert field.

In the charged sector, we can obtain a mass squared matrix in the basis $\left(\eta_{21}^{\pm},\eta_{22}^{\pm},\rho_1^{\pm},\rho_3^{\pm},\chi_2^{\pm}\right)$ of the form.
\begin{equation}
M^2_{\phi^{\pm}}=\left(
\begin{array}{ccccc}
\mu_{\eta_1}^2 +\left(\lambda-\lambda_4\right)v_{\eta}^2+\frac{\lambda_8}{2} v_{\rho}^2+\frac{\lambda_5}{2} v_{\chi}^2& 0 & 0 & 0 & 0 \\
 0 & \frac{A v_{\rho } v_{\chi }}{\sqrt{2} v_{\eta _2}} & \frac{A v_{\chi }}{\sqrt{2}} & 0 &
   0 \\
 0 & \frac{A v_{\chi }}{\sqrt{2}} & \frac{A v_{\chi } v_{\eta _2}}{\sqrt{2} v_{\rho }} & 0 &
   0 \\
 0 & 0 & 0 & \frac{A v_{\chi } v_{\eta _2}}{\sqrt{2} v_{\rho }} & \frac{A v_{\eta
   _2}}{\sqrt{2}} \\
 0 & 0 & 0 & \frac{A v_{\eta _2}}{\sqrt{2}} & \frac{A v_{\rho } v_{\eta _2}}{\sqrt{2}
   v_{\chi }} \\
\end{array}
\right)
\end{equation}

In sector CP odd, the square matrices in the pseudoscalar neutral basis $\left(\zeta_{\eta_1},\zeta_{\eta_2},\zeta_{\rho},\zeta_{\chi}\right)$
\begin{equation}
M^2_{\zeta}=\left(
\begin{array}{cccc}
\mu _{\eta _1}^2+\left(\lambda _2-2 \lambda _3-\lambda _4\right) v_{\eta _2}^2+\frac{\lambda_8}{2}v_{\rho}^2+\frac{\lambda_5}{2}v_{\chi}^2 & 0 & 0 & 0 \\
 0 & \frac{A v_{\rho } v_{\chi }}{\sqrt{2} v_{\eta
   _2}} & \frac{A v_{\chi }}{\sqrt{2}} & \frac{A
   v_{\rho }}{\sqrt{2}} \\
 0 & \frac{A v_{\chi }}{\sqrt{2}} & \frac{A
   v_{\chi } v_{\eta _2}}{\sqrt{2} v_{\rho }} &
   \frac{A v_{\eta _2}}{\sqrt{2}} \\
 0 & \frac{A v_{\rho }}{\sqrt{2}} & \frac{A
   v_{\eta _2}}{\sqrt{2}} & \frac{A v_{\rho }
   v_{\eta _2}}{\sqrt{2} v_{\chi }} \\
\end{array}
\right)
\end{equation}
and neutral scalar complex at the basis $\left(\text{Im}\eta_{31}^0,\text{Im}\eta_{32}^0,\text{Im}\chi_{1}^0 \right)$
\begin{equation}
M_{\phi_{\text{Im}}}^2=\left(
\begin{array}{ccc}
 2 \mu _{\eta _1}^2+2 \left(\lambda _2-\lambda
   _4\right) v_{\eta _2}^2+\lambda _8 v_{\rho
   }^2+\left(\lambda _5-\lambda _6\right) v_{\chi
   }^2 & 0 & 0 \\
 0 &
   \frac{\sqrt{2} A v_{\rho
   }v_{\chi }}{v_{\eta _2}}-\lambda _6 v_{\chi }^2&
   \lambda _6 v_{\eta _2} v_{\chi }-\sqrt{2} A
   v_{\rho } \\
 0 & \lambda _6 v_{\eta _2} v_{\chi }-\sqrt{2} A
   v_{\rho } & \frac{\sqrt{2} A
    v_{\eta _2} v_{\rho }}{v_{\chi }}-\lambda _6 v_{\eta
   _2}^2 \\
\end{array}
\right)
\end{equation}

In the sector CP even, the square mass matrices in the scalar neutral basis $\left(\xi_{\eta_1},\xi_{\eta_2},\xi_{\rho},\xi_{\chi}\right) $
\begin{equation}
M^2_\xi=\left(
\begin{array}{cccc}
\mu _{\eta _1}^2+\left(\lambda _2+\lambda _4\right)
   v_{\eta _2}^2+\frac{\lambda_8}{2}v_{\rho}^2+\frac{\lambda _5}{2} v_{\chi}^2 & 0 & 0 & 0 \\
 0 &
 \frac{A v_{\rho } v_{\chi }}{\sqrt{2} v_{\eta _2}^2}
 +2\left(\lambda_2+\lambda_4 \right)v_{\eta _2}^2
 & \lambda _8 v_{\eta _2}
   v_{\rho }-\frac{A v_{\chi }}{\sqrt{2}} &
   \lambda _5 v_{\eta _2} v_{\chi }-\frac{A
   v_{\rho }}{\sqrt{2}} \\
 0 & \lambda _8 v_{\eta _2} v_{\rho }-\frac{A
   v_{\chi }}{\sqrt{2}} & \frac{A v_{\eta _2}
   v_{\chi }}{\sqrt{2} v_{\rho }}+2 \lambda _7
   v_{\rho }^2 & \lambda _9 v_{\rho } v_{\chi
   }-\frac{A v_{\eta _2}}{\sqrt{2}} \\
 0 & \lambda _5 v_{\eta _2} v_{\chi }-\frac{A
   v_{\rho }}{\sqrt{2}} & \lambda _9 v_{\rho }
   v_{\chi }-\frac{A v_{\eta _2}}{\sqrt{2}} &
   \frac{A v_{\eta _2} v_{\rho }}{\sqrt{2} v_{\chi
   }}+2 \lambda _1 v_{\chi }^2 \\
\end{array}
\right)
\end{equation}
and neutral scalar complex at the basis $\left(\text{Re}\eta_{31}^0,\text{Re}\eta_{32}^0,\text{Re}\chi_{1}^0 \right)$
\begin{equation}
M_{\phi_{\text{Re}}}^2=\left(
\begin{array}{ccc}
 2 \mu _{\eta _1}^2+2 \left(\lambda _2-\lambda
   _4\right) v_{\eta _2}^2+\lambda _8 v_{\rho
   }^2+\left(\lambda _5+\lambda _6\right) v_{\chi
   }^2 & 0 & 0 \\
 0 &  \frac{\sqrt{2} A v_{\rho
   }v_{\chi }}{v_{\eta _2}}+\lambda _6 v_{\chi }^2 &
   \sqrt{2} A v_{\rho }+\lambda _6 v_{\eta _2}
   v_{\chi } \\
 0 & \sqrt{2} A v_{\rho }+\lambda _6 v_{\eta _2}
   v_{\chi } &  \frac{\sqrt{2} A
   v_{\eta _2}v_{\rho }}{v_{\chi }}+\lambda _6 v_{\eta
   _2}^2 \\
\end{array}
\right)
\end{equation}

\bibliographystyle{utphys}
\bibliography{Biblio331.bib}

\end{document}